\documentclass[twocolumn,pre,aps,showpacs,showkeys,amsmath]{revtex4}
\usepackage{graphicx}

\begin{document}

\title{Effect of Interpopulation Spike-Timing-Dependent Plasticity on Synchronized Rhythms in Neuronal Networks with Inhibitory and Excitatory Populations}
\author{Sang-Yoon Kim}
\email{sykim@icn.re.kr}
\author{Woochang Lim}
\email{wclim@icn.re.kr}
\affiliation{Institute for Computational Neuroscience and Department of Science Education, Daegu National University of Education, Daegu 42411, Korea}

\begin{abstract}
We consider a two-population network consisting of both inhibitory (I) interneurons and excitatory (E) pyramidal cells.
This I-E neuronal network has adaptive dynamic I to E and E to I interpopulation synaptic strengths, governed by interpopulation spike-timing-dependent plasticity (STDP). In previous works without STDPs, fast sparsely synchronized rhythms, related to diverse cognitive functions, were found to appear in a range of noise intensity $D$ for static synaptic strengths. Here, by varying $D$, we investigate the effect of interpopulation STDPs on fast sparsely synchronized rhythms that emerge in both the I- and the E-populations. Depending on values of $D$, long-term potentiation (LTP) and long-term depression (LTD) for population-averaged values of saturated interpopulation synaptic strengths are found to occur. Then, the degree of fast sparse synchronization varies due to effects of LTP and LTD. In a broad region of intermediate $D$, the degree of good synchronization (with higher synchronization degree) becomes decreased, while in a region of large $D$, the degree of bad synchronization (with lower synchronization degree) gets increased. Consequently, in each I- or E-population, the synchronization degree becomes nearly the same in a wide range of $D$ (including both the intermediate and the large $D$ regions). This kind of ``equalization effect'' is found to occur via cooperative interplay between the average occupation and pacing degrees of spikes (i.e., the average fraction of firing neurons and the average degree of phase coherence between spikes in each synchronized stripe of spikes in the raster plot of spikes) in fast sparsely synchronized rhythms. Finally, emergences of LTP and LTD of interpopulation synaptic strengths (leading to occurrence of equalization effect) are intensively investigated via a microscopic method based on the distributions of time delays between the pre- and the post-synaptic spike times.
\end{abstract}


\maketitle

\section{Introduction}
\label{sec:INT}
Recently, much attention has been paid to brain rhythms that emerge via population synchronization between individual firings in neuronal networks \cite{Buz1,TW,Rhythm1,Rhythm2,Rhythm3,Rhythm4,Rhythm5,Rhythm6,Rhythm7,Rhythm8,Rhythm9,Rhythm10,Rhythm11,Rhythm12,Rhythm13}.
In particular, we are concerned about fast sparsely synchronized rhythms, associated with diverse cognitive functions (e.g., multisensory feature binding, selective attention, and memory formation) \cite{W_Review}.
Fast sparsely synchronous oscillations [e.g., gamma rhythm (30-80 Hz) during awake behaving states and rapid eye movement sleep] have been observed in local field potential recordings, while at the cellular level individual neuronal recordings have been found to exhibit stochastic and intermittent spike discharges like Geiger counters at much lower rates than the population oscillation frequency \cite{FSS1,FSS2,FSS3,FSS4}. Hence, single-cell firing activity differs distinctly from the population oscillatory behavior. These fast sparsely synchronized rhythms are in contrast to fully synchronized rhythms where individual neurons fire regularly at the population oscillation frequency like clocks.

Diverse states (which exhibit synchronous or asynchronous population activity and regular or irregular single-cell activity) appear in the real brain.
In the case of asynchronous irregular state, recordings of local field potentials in the neocortex and the hippocampus in vivo do not exhibit prominent field oscillations, along with irregular firings of single cells at low frequencies (as shown in their Poisson-like histograms of interspike intervals \cite{AI1,AI2,AI3,AI4}). Hence, these asynchronous irregular states show stationary global activity and irregular single-cell firings with low frequencies \cite{AI5,AI6,Sparse1}. On the other hand, in the case of synchronous irregular state, prominent oscillations of local field potentials (corresponding to gamma rhythms) were observed in the hippocampus, the neocortex, the cerebellum, and the olfactory system \cite{FSS1,FSS2,FSS3,FSS4,Gamma1,Gamma2,Gamma3,Gamma4,Gamma5,Gamma6,Gamma7,Gamma8,Gamma9,Gamma10,Gamma11,Gamma12,Gamma13,Gamma14,Gamma15,Gamma16}.
We note that, even when recorded local field potentials exhibit fast synchronous oscillations, spike trains of single cells are still highly irregular and sparse
\cite{FSS1,FSS2,FSS3,FSS4}. For example, Csicsvari et al. \cite{FSS1} observed that hippocampal pyramidal cells and interneurons fire irregularly at lower rates ($\sim$ 1.5 Hz for pyramidal cells and $\sim$ 15 Hz for interneurons) than the population frequency of global gamma oscillation.
In this work, we are concerned about such fast sparsely synchronized rhythms which exhibit oscillatory global activity and stochastic and
sparse single-cell firings.

Fast sparse synchronization was found to emerge under balance between strong external noise and strong recurrent inhibition in single-population networks of purely inhibitory interneurons and also in two-population networks of both inhibitory interneurons and excitatory pyramidal cells \cite{W_Review,Sparse1,Sparse2,Sparse3,Sparse4,Sparse5,Sparse6}.
In neuronal networks, architecture of synaptic connections has been found to have complex topology which is neither regular nor completely random \cite{CN1,CN2,CN3,CN4,CN5,CN6,CN7,Sporns}. In recent works \cite{FSS-SWN,FSS-SFN,FSS-CSWN}, we studied the effects of network architecture on emergence of fast sparse synchronization in small-world, scale-free, and clustered small-world complex networks with sparse connections, consisting of inhibitory interneurons.
Thus, fast sparsely synchronized rhythms were found to appear, independently of network structure. In these works, synaptic coupling strengths were static.

However, in real brains synaptic strengths may vary for adjustment to the environment. Thus, synaptic strengths may be potentiated \cite{LTP1,LTP2,LTP3} or depressed \cite{LTD1,LTD2,LTD3,LTD4}. This synaptic plasticity provides the basis for learning, memory, and development \cite{Abbott1}. Here, we consider spike-timing-dependent plasticity (STDP) for the synaptic plasticity \cite{STDP1,STDP2,STDP3,STDP4,STDP5,STDP6,STDP7,STDP8}. For the STDP, the synaptic strengths change through an update rule depending on the relative time difference between the pre- and the post-synaptic spike times. Recently, effects of STDP on diverse types of synchronization in populations of coupled neurons were studied in various aspects \cite{Tass1,Tass2,Brazil1,Brazil2,Brazil3,SBS,SSS,BS-iSTDP}. Particularly, effects of inhibitory STDP (at inhibitory to inhibitory synapses) on fast sparse synchronization have been investigated in small-world networks of inhibitory fast spiking interneurons \cite{FSS-iSTDP}.

Synaptic plasticity at excitatory and inhibitory synapses is of great interest because it controls the efficacy of potential computational functions of excitation and inhibition. Studies of synaptic plasticity have been mainly focused on excitatory synapses between pyramidal cells, since excitatory-to-excitatory (E to E) synapses are most prevalent in the cortex and they form a relatively homogeneous population \cite{EtoE1,EtoE2,EtoE3,EtoE4,EtoE5,EtoE6,EtoE7}.
A Hebbian time window was used for the excitatory STDP (eSTDP) update rule \cite{STDP1,STDP2,STDP3,STDP4,STDP5,STDP6,STDP7,STDP8}. When a pre-synaptic spike precedes (follows) a post-synaptic spike, long-term potentiation (LTP) [long-term depression (LTD)] occurs. In contrast, synaptic plasticity at inhibitory synapses has attracted less attention mainly due to experimental obstacles and diversity of interneurons \cite{iSTDP1,iSTDP2,iSTDP3,iSTDP4,iSTDP5}.
With the advent of fluorescent labeling and optical manipulation of neurons according to their genetic types \cite{iExpM1,iExpM2}, inhibitory synaptic plasticity has also begun to be focused. Particularly, studies on inhibitory STDP (iSTDP) at inhibitory-to-excitatory (I to E) synapses  have been much made. Thus, iSTDP has been found to be diverse and cell-specific \cite{iSTDP1,iSTDP2,iSTDP3,iSTDP4,iSTDP5,iSTDP6,iSTDP7,iSTDP8,iSTDP9,iSTDP10,iSTDP11,iSTDP12}.

We are concerned about fast sparsely synchronized rhythms, related to diverse cognitive functions such as feature integration, selective attention, and memory formation \cite{W_Review} [e.g., gamma rhythm (30-80 Hz) during awake behaving states and rapid eye movement sleep]. They appear independently of network architecture \cite{FSS-SWN,FSS-SFN,FSS-CSWN}. Here, we consider clustered small-world networks with both inhibitory (I) and excitatory (E) populations. The inhibitory small-world network consists of fast spiking interneurons and the excitatory small-world network is composed of regular spiking pyramidal cells. We assume that random uniform connections are made between the I- and the E-populations. In the case that specific connectivity rule between the two populations is not known, it would be reasonable to assume random uniform connectivity. This is the same logic as the random matrix theory for studying statistics of energy levels in nuclear physics \cite{RMT}. For the random matrix theory, the matrix elements of the Hamiltonian are assumed to be random variables because of lack of knowledge on the matrix elements.

By taking into consideration interpopulation STDPs between the I- and E-populations, we investigate their effects on diverse properties of population and individual behaviors of fast sparsely synchronized rhythms by varying the noise intensity $D$ in the combined case of both I to E iSTDP and E to I eSTDP.
A time-delayed Hebbian time window is employed for the I to E iSTDP update rule. Such time-delayed Hebbian time window was found experimentally at inhibitory synapses onto principal excitatory stellate cells in the superficial layer II of the entorhinal cortex of rat \cite{ItoETW1}. On the other hand, an anti-Hebbian time window is used for the E to I eSTDP update rule. This type of anti-Hebbian time window was experimentally found at excitatory synapses onto the GABAergic Purkinje-like cell in electrosensory lobe of electric fish \cite{EtoITW1}.
We note that our present work is in contrast to previous works on fast sparse synchronization where STDPs were not considered in most cases \cite{Sparse1,Sparse2,Sparse3,Sparse4,Sparse5,Sparse6} and only in one case \cite{FSS-iSTDP}, intrapopulation I to I iSTDP was considered in an inhibitory small-world network of fast spiking interneurons.

In the presence of interpopulation STDPs, interpopulation synaptic strengths $\{ J_{ij}^{(XY)} \}$ between the source $Y$-population and the target $X$-population are evolved into limit values $\{ {J_{ij}^{(XY)}}^* \}$ saturated over a time course governed by the learning rate of the STDP rule.
Depending on $D$, mean values $\langle {J_{ij}^{(XY)}}^* \rangle$ of saturated limit values are potentiated [long-term potentiation (LTP)] or depressed [long-term depression (LTD)], in comparison with the initial mean value $J_0^{(XY)}$. The degree of fast sparse synchronization changes because of the effects of LTP and LTD.
In the case of I to E iSTDP, LTP (LTD) disfavors (favors) fast sparse synchronization [i.e., LTP (LTD) tends to decrease (increase) the degree of fast sparse synchronization] due to increase (decrease) in the mean value of I to E synaptic inhibition. On the other hand, the roles of LTP and LTD are reversed in the case of E to I eSTDP. In this case, LTP (LTD) favors (disfavors) fast sparse synchronization [i.e., LTP (LTD) tends to increase (decrease) the degree of fast sparse synchronization] because of increase (decrease) in the mean value of E to I synaptic excitation.

Due to the effects of the mean (LTP or LTD), an ``equalization effect'' in interpopulation (both I to E and E to I) synaptic plasticity is found to emerge in a wide range of $D$ through cooperative interplay between the average occupation and pacing degrees of spikes (i.e., the average fraction of firing neurons and the average degree of phase coherence between spikes in each synchronized stripe of spikes in the raster plot of spikes) in fast sparsely synchronized rhythms. In a broad region of intermediate $D$, the degree of good synchronization (with higher synchronization degree) becomes decreased due to LTP (LTD) in the case of I to E iSTDP (E to I eSTDP). On the other hand, in a region of large $D$ the degree of bad synchronization (with lower synchronization degree) gets increased because of
LTD (LTP) in the case of I to E iSTDP (E to I eSTDP). Consequently, the degree of fast sparse synchronization becomes nearly the same in a wide range of $D$.

The degree of fast sparse synchronization is measured by employing the spiking measure \cite{RM}. Then, the equalization effect may be well visualized in the histograms of spiking measures (representing synchronization degree) in the absence and in the presence of interpopulation STDPs.
The standard deviation from the mean in the histogram in the case of interpopulation STDPs is much smaller than that in the case without STDP, which clearly shows emergence of the equalization effect. We also note that this kind of equalization effect in interpopulation synaptic plasticity is distinctly in contrast to the Matthew (bipolarization) effect in intrapopulation (I to I and E to E) synaptic plasticity where good (bad) synchronization gets better (worse) \cite{SSS,FSS-iSTDP}.

Emergences of LTP and LTD of interpopulation synaptic strengths (resulting in occurrence of equalization effect in interpopulation synaptic plasticity) are also investigated through a microscopic method based on the distributions of time delays $\{ \Delta t_{ij}^{(XY)} \}$ between the nearest spiking times of the post-synaptic neuron $i$ in the (target) $X$-population and the pre-synaptic neuron $j$ in the (source) $Y$-population. We follow time evolutions of normalized histograms $H(\Delta t_{ij}^{(XY)})$ in both cases of LTP and LTD. Because of the equalization effects, the two normalized histograms at the final (evolution) stage are nearly the same, which is in contrast to the case of intrapopulation STDPs where the two normalized histograms at the final stage are distinctly different due to the Matthew (bipolarization) effect \cite{SSS,FSS-iSTDP}.

This paper is organized as follows. In Sec.~\ref{sec:CSWN}, we describe clustered small-world networks composed of fast spiking interneurons (inhibitory small-world network) and regular spiking pyramidal cells (excitatory small-world network) with interpopulation STDPs. Then, in Sec.~\ref{sec:InterSTDP} the effect of interpopulation STDPs on fast sparse synchronization is investigated in the combined case of both I to E iSTDP and E to I eSTDP.
Finally, we give summary and discussion in Sec.~\ref{sec:SUM}.

\section{Clustered Small-World Networks Composed of Both I- and E-populations with Interpopulation Synaptic Plasticity}
\label{sec:CSWN}
In this section, we describe our clustered small-world networks consisting of both I- and E-populations with interpopulation synaptic plasticity.
A neural circuit in the brain cortex is composed of a few types of excitatory principal cells and diverse types of inhibitory interneurons. It is also known that interneurons make up about 20 percent of all cortical neurons, and exhibit diversity in their morphologies and functions \cite{Buz2}.
Here, we consider clustered small-world networks composed of both I- and E-populations. Each I(E)-population is modeled as a directed  Watts-Strogatz
small-world network, consisting of $N_I$ ($N_E$) fast spiking interneurons (regular spiking pyramidal cells) equidistantly placed on a one-dimensional ring of radius ${N_I}~(N_E) / 2 \pi$ ($N_I:N_E=1:4$), and random uniform connections with the probability $p_{inter}$ are made between the two inhibitory and excitatory small-world networks.

\begin{figure}
\includegraphics[width=0.7\columnwidth]{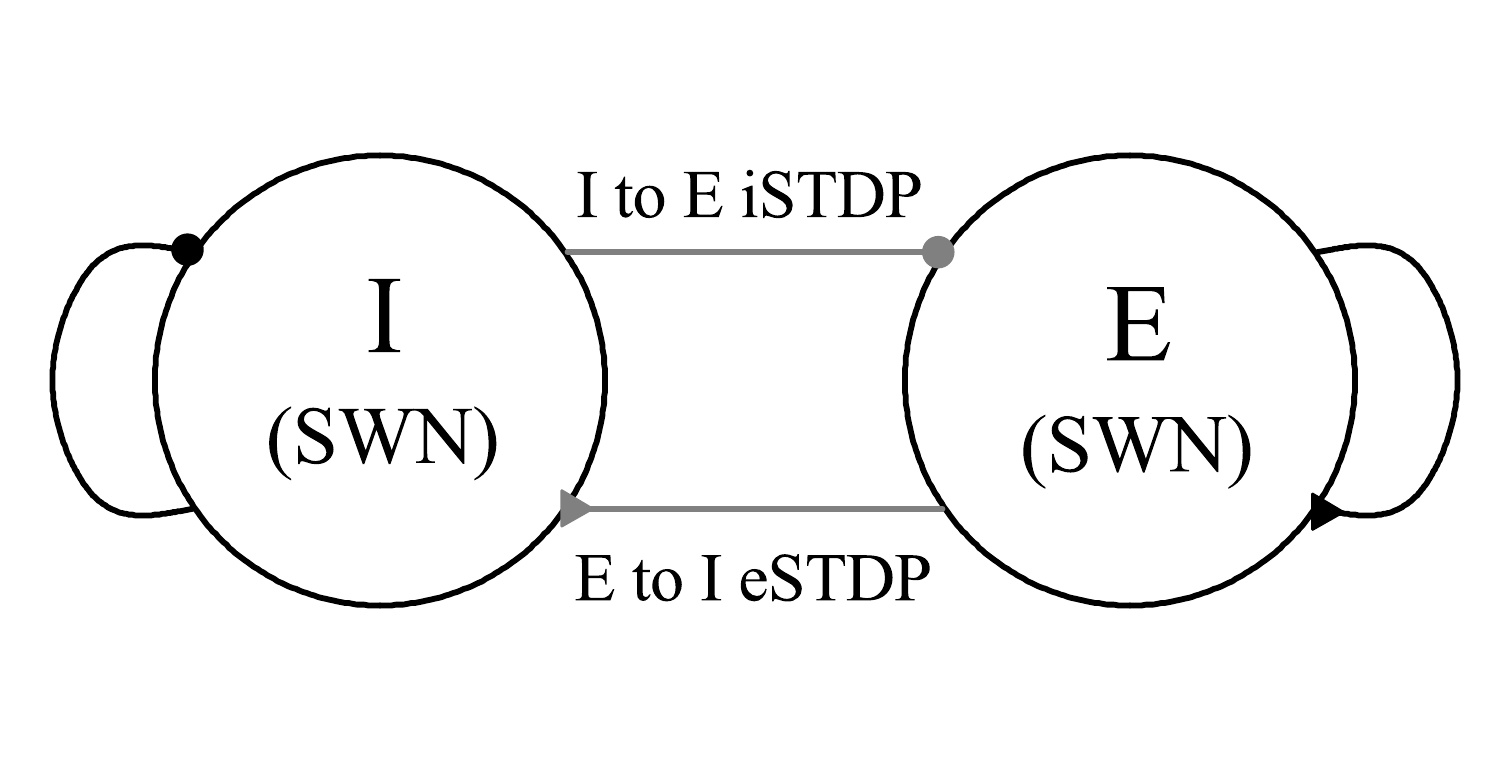}
\caption{Schematic representation of clustered small-world networks (SWNs) of the inhibitory (I) and the excitatory (E) populations with random interpopulation connections. Black curves with circle and triangle represent the I to I and the E to E intrapopulation connections, respectively. Gray lines with circle and triangle denote the I to E and the E to I interpopulation connections, respectively.}
\label{fig:CSWN}
\end{figure}

A schematic representation of the clustered small-world networks is shown in Fig.~\ref{fig:CSWN}.
The Watts-Strogatz inhibitory small-world network (excitatory small-world network) interpolates between a regular lattice with high clustering (corresponding to the case of $p_{wiring}=0$) and a random graph with short average path length (corresponding to the case of $p_{wiring}=1$) through random uniform rewiring with the probability $p_{wiring}$ \cite{SWN1,SWN2,SWN3}. For $p_{wiring}=0,$ we start with a directed regular ring lattice with $N_I$ ($N_E$) nodes where each node is coupled to its first $M_{syn}^{(I)}$ ($M_{syn}^{(E)}$) neighbors [$M_{syn}^{(I)}/2$ ($M_{syn}^{(E)}/2$) on either side] through outward synapses, and rewire each outward connection uniformly at random over the whole ring with the probability $p_{wiring}$ (without self-connections and duplicate connections).
Throughout the paper, we consider the case of $p_{wiring}=0.25$.
This kind of Watts-Strogatz small-world network model with predominantly local connections and rare long-range connections may be regarded as a cluster-friendly extension of the random network by reconciling the six degrees of separation (small-worldness) \cite{SDS1,SDS2} with the circle of friends (clustering).

As elements in the inhibitory small-world network (excitatory small-world network), we choose the Izhikevich inhibitory fast spiking interneuron (excitatory
regular spiking pyramidal cell) model which is not only biologically plausible, but also computationally efficient \cite{Izhi1,Izhi2,Izhi3,Izhi4}.
Unlike Hodgkin-Huxley-type conductance-based models, instead of matching neuronal electrophysiology, the Izhikevich model matches neuronal dynamics by tuning its parameters in the Izhikevich neuron model. The parameters $k$ and $b$ are related to the neuron's rheobase and input resistance, and $a,$ $c$, and $d$ are the recovery time constant, the after-spike reset value of $v$, and the after-spike jump value of $u$, respectively.

\begin{table*}
\caption{Parameter values used in our computations; units of the capacitance, the potential, the current, and the time are pF, mV, pA, and msec, respectively.}
\label{tab:Parm}
\begin{ruledtabular}
\begin{tabular}{lllllllll}
(1) & \multicolumn{8}{l}{Single Izhikevich Fast Spiking Interneurons \cite{Izhi3}} \\
&  $C_{I}=20$ & $v^{(I)}_{r}=-55$ & $v^{(I)}_{t}=-40$ & $v^{(I)}_{p}=25$ & $v^{(I)}_{b}=-55$ & & & \\
&  $k_{I}=1$ & $a_{I}=0.2$ & $b_{I}=0.025$ & $c_{I}=-45$ & $d_{I}=0$ & & & \\
\hline
(2) & \multicolumn{8}{l}{Single Izhikevich Regular Spiking Pyramidal Cells \cite{Izhi3}} \\
&  $C_{E}=100$ & $v^{(E)}_{r}=-60$ & $v^{(E)}_{t}=-40$ & $v^{(E)}_{p}=35$ & & & & \\
&  $k_{E}=0.7$ & $a_{E}=0.03$ & $b_{E}=-2$ & $c_{E}=-50$ & $d_{E}=100$ & & & \\
\hline
(3) & \multicolumn{8}{l}{Random External Excitatory Input to Each Izhikevich Fast Spiking Interneurons and Regular Spiking Pyramidal Cells} \\
& \multicolumn{3}{l}{${\overline{I_i}}^{(I)}={\overline{I_i}}^{(E)}= \overline{I_i}$; $\overline{I_i} \in [680, 720]$} & \multicolumn{5}{l}{$D_{I}=D_{E}=D$: Varying} \\
\hline
(4) & \multicolumn{8}{l}{Inhibitory Synapse Mediated by The GABA$_{\rm A}$ Neurotransmitter \cite{Sparse3}} \\
& I to I: & $\tau^{(II)}_l=1.5$ & $\tau^{(II)}_r=1.5$ & $\tau^{(II)}_d=8$ & $V^{(I)}_{syn}=-80$ & & & \\
& I to E: & $\tau^{(EI)}_l=1.5$ & $\tau^{(EI)}_r=1.5$ & $\tau^{(EI)}_d=8$ & & & & \\
\hline
(5) & \multicolumn{8}{l}{Excitatory Synapse Mediated by The AMPA Neurotransmitter \cite{Sparse3}} \\
& E to E: & $\tau^{(EE)}_l=1.5$ & $\tau^{(EE)}_r=0.4$ & $\tau^{(EE)}_d=2$ & $V^{(E)}_{syn}=0$ & & & \\
& E to I: & $\tau^{(IE)}_l=1.5$ & $\tau^{(IE)}_r=0.2$ & $\tau^{(IE)}_d=1$ & & & & \\
\hline
(6) & \multicolumn{8}{l}{Intra- and Inter-population Synaptic Connections between Neurons in The Clustered Watts-Strogatz Small-World}\\
& \multicolumn{8}{l}{Networks with Inhibitory and Excitatory Populations} \\
& \multicolumn{3}{l}{Intrapopulation Synaptic Connection:} & $N_{I}=600$ & $M^{(I)}_{syn}=40$ & $N_{E}=2400$ & $M^{(E)}_{syn}=160$ & $p_{wiring}=0.25$ \\
& \multicolumn{3}{l}{Interpopulation Synaptic Connection:} & $p_{inter}=1/15$ & & & & \\
& \multicolumn{2}{l}{Synaptic Strengthes:} & $J^{(II)}_{0}=1300$ & $J^{(EE)}_{0}=300$ & $J^{(EI)}_{0}=800$ & \multicolumn{3}{l}{$J^{(IE)}_{0}=487.5 ~~(=J^{(II)}_{0} J^{(EE)}_{0} / J^{(IE)}_{0})$} \\
& & & $\sigma_{0} = 5$ & \multicolumn{2}{l}{$J^{(EI)}_{ij} \in [0.0001, 2000]$} & \multicolumn{2}{l}{$J^{(IE)}_{ij} \in [0.0001, 2000]$} \\
\hline
(7) & \multicolumn{8}{l}{Delayed Hebbian I to E iSTDP Rule} \\
& $\delta = 0.1$ & $A_{+} = 0.4$ & $A_{-} = 0.35$ & $\tau_{+} = 2.6$ & $\tau_{-} = 2.8$ & & & \\
\hline
(8) & \multicolumn{8}{l}{Anti-Hebbian E to I eSTDP Rule} \\
& $\delta = 0.05$ & $A_{+} = 1.0$ & $A_{-} = 0.9$ & $\tau_{+} = 15.0$ & $\tau_{-} = 15.0$ & & & \\
\end{tabular}
\end{ruledtabular}
\end{table*}

Tuning the above parameters, the Izhikevich neuron model may produce 20 of the most prominent neuro-computational features of biological neurons \cite{Izhi1,Izhi2,Izhi3,Izhi4}.
In particular, the Izhikevich model is employed to reproduce the six most fundamental classes of firing patterns observed in the mammalian neocortex; (i) excitatory regular spiking pyramidal cells, (ii) inhibitory fast spiking interneurons, (iii) intrinsic bursting neurons, (iv) chattering neurons, (v) low-threshold spiking neurons, and (vi) late spiking neurons \cite{Izhi3}.
Here, we use the parameter values for the fast spiking interneurons and the regular spiking pyramidal cells in the layer 5 rat visual cortex, which are listed in the 1st ad the 2nd items of Table \ref{tab:Parm} (see the captions of Figs. 8.12 and 8.27 in \cite{Izhi3}).

The following equations (\ref{eq:PD1})-(\ref{eq:PD11}) govern population dynamics in the clustered small-world networks with the I- and the E-populations:
\begin{widetext}
\begin{eqnarray}
C_{I}\frac{dv_{i}^{(I)}}{dt} &=& k_{I} (v_i^{(I)} - v_r^{(I)}) (v_i^{(I)} - v_t^{(I)}) - u_i^{(I)} + {\overline{I_i}}^{(I)} +D_{I} \xi_{i}^{(I)} -I_{syn,i}^{(II)} -I_{syn,i}^{(IE)}, \label{eq:PD1} \\
\frac{du_i^{(I)}}{dt} &=& a_I \{ U(v_i^{(I)}) - u_i^{(I)} \},  \;\;\; i=1, \cdots, N_I, \label{eq:PD2} \\
C_{E}\frac{dv_{i}^{(E)}}{dt} &=& k_{E} (v_i^{(E)} - v_r^{(E)}) (v_i^{(E)} - v_t^{(E)}) - u_i^{(E)} +
{\overline{I_i}}^{(E)} +D_{E} \xi_{i}^{(E)} -I_{syn,i}^{(EE)} -I_{syn,i}^{(EI)}, \label{eq:PD3} \\
\frac{du_i^{(E)}}{dt} &=& a_E \{ U(v_i^{(E)}) - u_i^{(E)} \},  \;\;\; i=1, \cdots, N_E, \label{eq:PD4}
\end{eqnarray}
with the auxiliary after-spike resetting:
\begin{equation}
{\rm if~} v_i^{(X)} \geq v_p^{(X)},~ {\rm then~} v_i^{(X)} \leftarrow c_X ~ {\rm and~} u_i^{(X)} \leftarrow u_i^{(X)} + d_X,~~ (X=I ~\textrm{or}~ E)\label{eq:PD5}
\end{equation}
where
\begin{eqnarray}
U(v^{(I)}) &=& \left\{ \begin{array}{l} 0 {\rm ~for~} v^{(I)}<v_b^{(I)} \\ b_I (v^{(I)} - v_b^{(I)})^3 {\rm ~for~} v^{(I)} \ge v_b^{(I)} \end{array} \right. , \label{eq:PD6} \\
U(v^{(E)}) &=& b_E (v^{(E)} - v_b^{(E)}), \label{eq:PD7} \\
I_{syn,i}^{(XX)}(t) &=& \frac{1}{d_{in,i}^{intra}} \sum_{j=1 (j \ne i)}^{N_X} J_{ij}^{(XX)} w_{ij}^{(XX)} s_j^{(XX)}(t) (v_i^{(X)} - V_{syn}^{(X)}), \label{eq:PD8}\\
I_{syn,i}^{(XY)}(t) &=& \frac{1}{d_{in,i}^{inter}} \sum_{j=1}^{N_Y} J_{ij}^{(XY)} w_{ij}^{(XY)} s_j^{(XY)}(t) (v_i^{(X)} - V_{syn}^{(Y)}), \label{eq:PD9}\\
s_j^{(XY)}(t) &=& \sum_{f=1}^{F_j} E_{XY}(t-t_f^{(j)}-\tau_l^{(XY)}) ~(X=Y {\rm or} X \neq Y); \label{eq:PD10} \\
E_{XY}(t) &=& \frac{1}{\tau_d^{(XY)} - \tau_r^{(XY)}} (e^{-t/\tau_d^{(XY)}} - e^{-t/\tau_r^{(XY)}}) \Theta(t). \label{eq:PD11}
\end{eqnarray}
\end{widetext}

Here, the state of the $i$th neuron in the $X$-population ($X=I$ or $E$) at a time $t$ is characterized by two state variables: the membrane potential $v_i^{(X)}$ and the recovery current $u_i^{(X)}$. In Eq.~(\ref{eq:PD1}), $C_X$ is the membrane capacitance, $v_r^{(X)}$ is the resting membrane potential, and $v_t^{(X)}$ is the instantaneous threshold potential. After the potential reaches its apex (i.e., spike cutoff value) $v_p^{(X)}$, the membrane potential and the recovery variable are reset according to Eq.~(\ref{eq:PD5}). The units of the capacitance $C_X$, the potential $v^{(X)}$, the current $u^{(X)}$ and the time $t$ are pF, mV, pA, and msec, respectively. All these parameter values used in our computations are listed in Table \ref{tab:Parm}. More details on the random external input, the synaptic currents and plasticity, and the numerical method for integration of the governing equations are given in the following subsections.

\subsection{Random External Excitatory Input to Each Izhikevich Fast Spiking Interneuron and Regular Spiking Pyramidal Cell}
\label{subsec:Ext}
Each neuron in the $X$-population ($X=I$ or $E$) receives stochastic external excitatory input $I_{ext,i}^{(X)}$ from other brain regions, not included in the network (i.e., corresponding to background excitatory input)
\cite{Sparse1,Sparse2,Sparse3,Sparse4}. Then, $I_{ext,i}^{(X)}$ may be modeled in terms of its time-averaged constant $\overline{I_{i}}^{(X)}$ and an independent Gaussian white noise $\xi_i^{(X)}$ (i.e., corresponding to fluctuation of
$I_{ext,i}^{(X)}$ from its mean) [see the 3rd and the 4th terms in Eqs.~(\ref{eq:PD1}) and (\ref{eq:PD3})]
satisfying $\langle \xi_i^{(X)}(t) \rangle =0$ and $\langle \xi_i^{(X)}(t)~\xi_j^{(X)}(t') \rangle = \delta_{ij}~\delta(t-t')$, where $\langle\cdots\rangle$ denotes the ensemble average. The intensity of the noise $\xi_i^{(X)}$ is controlled by using the parameter $D_X$. For simplicity, we consider the case of $\overline{I_{i}}^{(I)} = \overline{I_{i}}^{(E)} = \overline{I_{i}}$ and $D_I=D_E=D$.

\begin{figure}
\includegraphics[width=0.7\columnwidth]{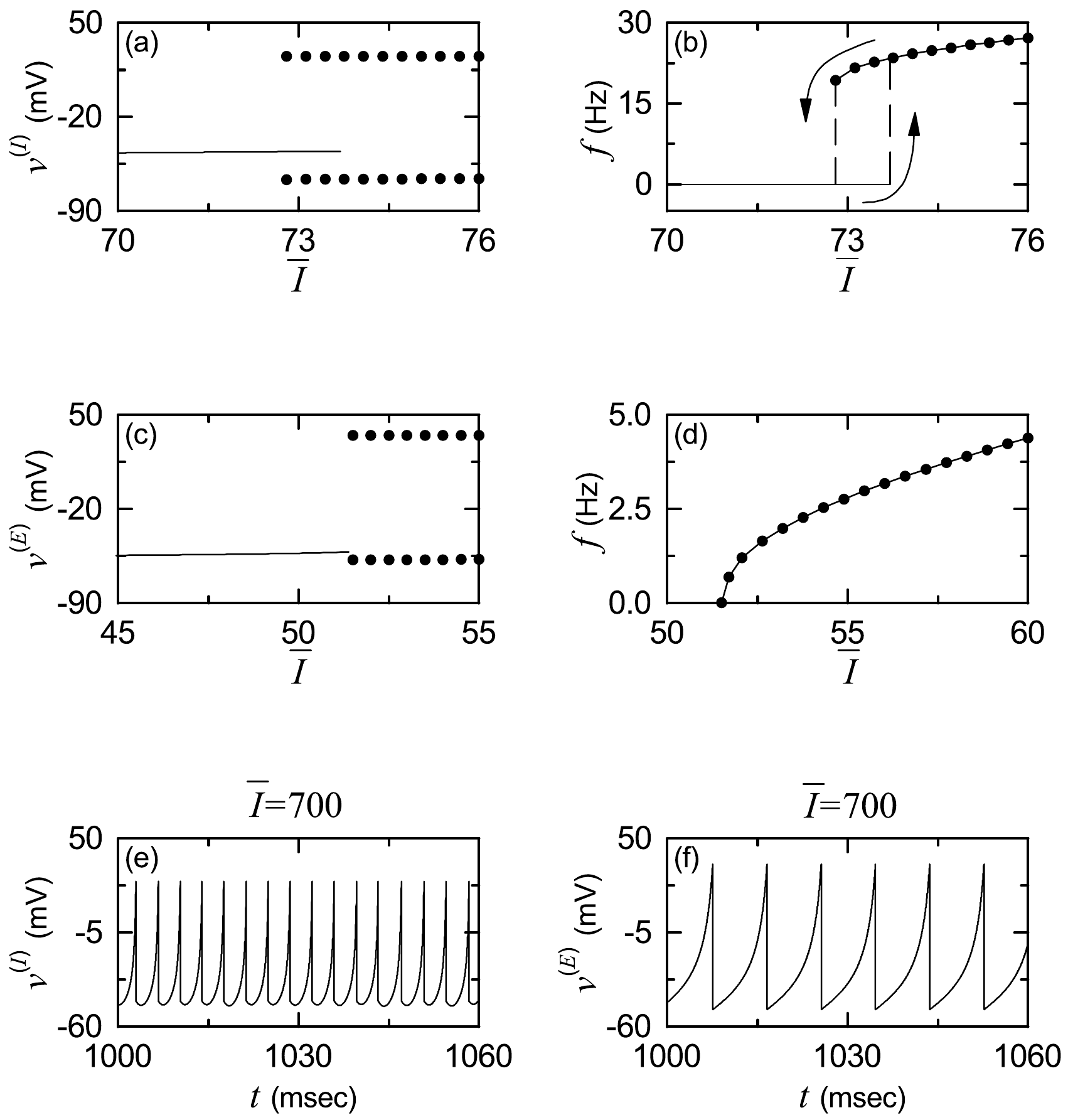}
\caption{Single Izhikevich fast spiking interneuron for $D=0$: (a) bifurcation diagram (i.e., plot of $v^{(I)}$ versus the time-averaged constant $\overline{I}$ of the external input $I_{ext}$) and (b) plot of the mean firing rate $f$  versus $\overline {I}$. Single Izhikevich regular spiking pyramidal cell for $D=0$: (c) bifurcation diagram (i.e, plot of $v^{(E)}$ versus $\overline{I}$) and (d) plot of $f$ versus $\overline {I}$. In (a) and (c), solid lines denote stable equilibrium points, and solid circles represent maximum and minimum values of the membrane potential $v^{(X)}$ ($X=I$ or $E$) for the spiking states. Time series of the membrane potential $v^{(X)}$ of (e) the Izhikevich fast spiking interneuron and (f) the Izhikevich regular spiking pyramidal cell for $\overline {I} =700$.
}
\label{fig:Single}
\end{figure}

Figure \ref{fig:Single} shows spiking transitions for both the single Izhikevich fast spiking interneuron and regular spiking pyramidal cell
in the absence of noise (i.e., $D=0$). The fast spiking interneuron exhibits a jump from a resting state to a spiking state via subcritical Hopf bifurcation for ${\overline{I}}_h^* \simeq 73.7$ by absorbing an unstable limit cycle born via a fold limit cycle bifurcation for ${\overline{I}}_l^* \simeq 72.8$ [see Fig.~\ref{fig:Single}(a)] \cite{Izhi3}. Hence, the fast spiking interneuron shows type-II excitability because it begins to fire with a non-zero frequency, as shown in Fig.~\ref{fig:Single}(b) \cite{Ex1,Ex2}. Throughout this paper, we consider a suprathreshold case such that the value of $\overline{I_i}$ is chosen via uniform random sampling in the range of [680,720], as shown in the 3rd item of Table \ref{tab:Parm}. At the middle value of $\overline{I}=700$, the membrane potential $v^{(I)}$ oscillates very fast with a mean firing rate $f \simeq 271$ Hz [see Fig.~\ref{fig:Single}(e)]. On the other hand, the regular spiking pyramidal cell shows a continuous transition from a resting state to a spiking state through a saddle-node bifurcation on an invariant circle for ${\overline{I}}^* \simeq 51.5$, as shown in Fig.~\ref{fig:Single}(c) \cite{Izhi3}. Hence, the regular spiking pyramidal cell exhibits type-I excitability because its frequency $f$ increases continuously from 0 [see Fig.~\ref{fig:Single}(d)]. For $\overline {I} =700$, the membrane potential $v^{(E)}$ oscillates with $f \simeq 111$ Hz, as shown in Fig.~\ref{fig:Single}(f). Hence, $v^{(I)}(t)$ (of the fast spiking interneuron) oscillates about 2.4 times as fast as $v^{(E)}(t)$
(of the regular spiking pyramidal cell) when $\overline {I} =700$.

\subsection{Synaptic Currents and Plasticity}
\label{subsec:Syn}
Here, we choose the numbers of fast spiking interneurons and regular spiking pyramidal cells as $N_I=600$ and $N_E=2400,$ respectively which
satisfy the 1: 4 ratio (i.e, $N_I:N_E = 1:4$).
The last two terms in Eq.~(\ref{eq:PD1}) represent synaptic couplings of fast spiking interneurons in the I-population with $N_I=600$.
$I_{syn,i}^{(II)}(t)$ and $I_{syn,i}^{(IE)}(t)$ in Eqs.~(\ref{eq:PD8}) and (\ref{eq:PD9}) denote intrapopulation I to I synaptic current and interpopulation E to I synaptic current injected into the fast spiking interneuron $i$, respectively, and
$V_{syn}^{(I)}$ [$V_{syn}^{(E)}$] is the synaptic reversal potential for the inhibitory (excitatory) synapse.
Similarly, regular spiking pyramidal cells in the E-population with $N_E=2400$ also have two types of synaptic couplings [see the last two terms in Eq.~(\ref{eq:PD3})]. In this case, $I_{syn,i}^{(EE)}(t)$ and $I_{syn,i}^{(EI)}(t)$ in Eqs.~(\ref{eq:PD8}) and (\ref{eq:PD9}) represent intrapopulation E to E synaptic current and interpopulation I to E synaptic current injected into the regular spiking pyramidal cell $i$, respectively.

The intrapopulation synaptic connectivity in the $X$-population ($X=I$ or $E$) is given by the connection weight matrix $W^{(XX)}$ (=$\{ w_{ij}^{(XX)} \}$) where  $w_{ij}^{(XX)}=1$ if the neuron $j$ is pre-synaptic to the neuron $i$; otherwise, $w_{ij}^{(XX)}=0$. Here, the intrapopulation synaptic connection is modeled in terms of the Watts-Strogatz small-world network.
Then, the intrapopulation in-degree of the neuron $i$, $d_{in,i}^{intra}$ (i.e., the number of intrapopulation synaptic inputs to the neuron $i$) is given by $d_{in,i}^{intra} = \sum_{j=1 (j \ne i)}^{N_X} w_{ij}^{(XX)}$. In this case, the average number of intrapopulation synaptic inputs per neuron is given by $M_{syn}^{(X)} = \frac{1}{N_X} \sum_{i=1}^{N_X} d_{in,i}^{intra}$. 
Throughout the paper, we consider a sparsely connected case of $M_{syn}^{(I)} = 40$ in the I-population with $N_I=600$. In this case, the sparseness degree for the synaptic inputs to each interneuron may be given by $M_{syn}^{(I)} / N_I = 1/15$. In the E-population with $N_E=2400$, we also consider the case with the same sparseness degree (i.e., 1/15), which leads to $M_{syn}^{(E)} = 160$. These values of $M_{syn}^{(I)} =40$ and $M_{syn}^{(E)} = 160$ are shown in the 6th item of Table \ref{tab:Parm}.

Next, we consider interpopulation synaptic couplings. The interpopulation synaptic connectivity from the source $Y$-population to the target $X$-population is given by the connection weight matrix $W^{(XY)}$ (=$\{ w_{ij}^{(XY)} \}$) where  $w_{ij}^{(XY)}=1$ if the neuron $j$ in the source $Y$-population is pre-synaptic to the neuron $i$ in the target $X$-population; otherwise, $w_{ij}^{(XY)}=0$. Random uniform connections are made with the probability $p_{inter}$ between the two I- and E-populations. Here, we consider the case of $p_{inter}=1/15$ which is the same as the sparseness degree for the intrapolulation connections. Then, the average number of E to I synaptic inputs per each fast spiking interneuron and I to E synaptic inputs per each regular spiking pyramidal cell are 160 and 40, respectively.

We consider synapses from the $Y$ source population to the $X$ target population. The post-synaptic ion channels are opened due to the binding of neurotransmitters (emitted from the $Y$ source population) to receptors in the $X$ target population. The fraction of open ion channels at time $t$ is
denoted by $s^{(XY)}(t)$. The time course of $s_j^{(XY)}(t)$ of the neuron $j$ in the source $Y$-population is given by a sum of delayed double-exponential functions $E_{XY}(t-t_f^{(j)}-\tau_l^{(XY)})$ [see Eq.~(\ref{eq:PD10})], where $\tau_l^{(XY)}$ is the synaptic delay for the $Y$ to $X$ synapse, and $t_f^{(j)}$ and $F_j$ are the $f$th spiking time and the total number of spikes of the $j$th neuron in the $Y$-population at time $t$, respectively. Here, $E_{XY}(t)$ in Eq.~(\ref{eq:PD11}) [which corresponds to contribution of a pre-synaptic spike occurring at time $0$ to $s_j^{(XY)}(t)$ in the absence of synaptic delay] is controlled by the two synaptic time constants: synaptic rise time $\tau_r^{(XY)}$ and decay time $\tau_d^{(XY)}$, and $\Theta(t)$ is the Heaviside step function: $\Theta(t)=1$ for $t \geq 0$ and 0 for $t <0$. For the inhibitory GABAergic synapse (involving the $\rm{GABA_A}$ receptors), the values of $\tau_l^{(XI)}$, $\tau_r^{(XI)}$, $\tau_d^{(XI)}$, and $V_{syn}^{(I)}$ ($X=I$ or $E$) are listed in the 4th item of Table \ref{tab:Parm} \cite{Sparse3}. For the excitatory AMPA synapse (involving the AMPA receptors), the values of $\tau_l^{(XE)}$, $\tau_r^{(XE)}$, $\tau_d^{(XE)}$, and $V_{syn}^{(E)}$ ($X=E$ or $I$) are given in the 5th item of Table \ref{tab:Parm} \cite{Sparse3}.

The coupling strength of the synapse from the pre-synaptic neuron $j$ in the source $Y$-population to the post-synaptic neuron $i$ in the target $X$-population is $J_{ij}^{(XY)}$; for the intrapopulation synaptic coupling $X=Y$,
while for the interpopulation synaptic coupling, $X \neq Y$.
Initial synaptic strengths are normally distributed with the mean $J_0^{(XY)}$ and the standard deviation $\sigma_0~(=5)$.
Here, $J_0^{(II)}=1300,$ $J_0^{(EE)}=300$, $J_0^{(EI)}=800$, $J_0^{(IE)}=487.5$ (=$J_0^{(II)}~J_0^{(EE)} / J_0^{(EI)}$)
(see the 6th item of Table \ref{tab:Parm}). In this initial case, the E-I ratio (given by the ratio of average excitatory to inhibitory synaptic strengths) is the same in both fast spiking interneurons and regular spiking pyramidal cells [i.e., $J_0^{(EE)} / J_0^{(EI)}$ (E-population) =
$J_0^{(IE)} / J_0^{(II)}$ (I-population)] \cite{Sparse3,Sparse4,Sparse6}. Hereafter, this will be called the ``E-I ratio balance," because the E-I ratios in both E- and I-populations are balanced. In our previous works \cite{SSS,FSS-iSTDP}, we studied the effect of intrapopulation (E to E and I to I) synaptic plasticity on synchronized rhythms, and the Matthew (bipolarization) effect where good (bad) synchronization becomes better (worse) has thus been found. Here, we restrict our attention only to the interpopulation (I to E and E to I) synaptic plasticity. Thus, intrapopulation synaptic strengths are static in the present study.

For the interpopulation synaptic strengths $\{ J_{ij}^{(XY)} \},$ we consider a multiplicative STDP (dependent on states) \cite{Multi,Tass2,FSS-iSTDP}. To avoid unbounded growth and elimination of synaptic connections, we set a range with the upper and the lower bounds:
$J_{ij}^{(XY)} \in [J_l, J_h]$, where $J_l=0.0001$ and $J_h=2000$.
With increasing time $t$, synaptic strength for each interpopulation synapse is updated with a nearest-spike pair-based STDP rule \cite{SS}:
\begin{equation}
J_{ij}^{(XY)} \rightarrow J_{ij}^{(XY)} + \delta (J^*-J_{ij}^{(XY)})~|\Delta J_{ij}^{(XY)}(\Delta t_{ij}^{(XY)})|,
\label{eq:MSTDP}
\end{equation}
where $J^*=$ $J_h~(J_l)$ for the LTP (LTD) and $\Delta J_{ij}^{(XY)}(\Delta t_{ij}^{(XY)})$ is the synaptic modification depending on the relative time difference $\Delta t_{ij}^{(XY)}$ $(=t_i^{(post,X)} - t_j^{(pre,Y)})$ between the nearest spike times of the post-synaptic neuron $i$ in the target $X$-population and the pre-synaptic neuron $j$
in the source $Y$-population. The values of the update rate $\delta$ for the I to E iSTDP and the E to I eSTDP are
0.1 and 0.05, respectively (see the 7th and the 8th items of Table \ref{tab:Parm})

For the I to E iSTDP, we use a time-delayed Hebbian time window for the synaptic modification $\Delta J_{ij}^{(EI)}(\Delta t_{ij}^{(EI)})$ \cite{ItoETW1,ItoETW2,Brazil2}:
\begin{equation}
  \Delta J_{ij}^{(EI)}(\Delta t_{ij}^{(EI)}) = \left\{
         \begin{array}{l}
           E_+(\Delta t_{ij}^{(EI)})~ {\Delta t_{ij}^{(EI)}}^\beta  ~{\rm for}~\Delta t_{ij}^{(EI)} \geq 0 \\
           E_-(\Delta t_{ij}^{(EI)})~ {\Delta t_{ij}^{(EI)}}^\beta  ~{\rm for}~\Delta t_{ij}^{(EI)} < 0
         \end{array}
         \right. .
\label{eq:ItoETW}
\end{equation}
Here, $E_+(\Delta t_{ij}^{(EI)})$ and $E_-(\Delta t_{ij}^{(EI)})$ are Hebbian exponential functions used in the case of E to E eSTDP \cite{STDP1,SSS}:
\begin{eqnarray}
E_+(\Delta t_{ij}^{(EI)}) &=& A_+~ N_+~ e^{-\Delta t_{ij}^{(EI)} / \tau_{+}}~{\rm and}~ \nonumber \\
E_-(\Delta t_{ij}^{(EI)}) &=& A_-~ N_-~ e^{\Delta t_{ij}^{(EI)} / \tau_{-}},
\label{eq:Exp}
\end{eqnarray}
where $N_+ = \frac {e^\beta} {\beta^\beta~ \tau_+^\beta}$, $N_- = \frac {e^\beta} {\beta^\beta~ \tau_-^\beta}$,
$\beta=10$, $A_+=0.4$, $A_-=0.35$, $\tau_+=2.6$ msec, and $\tau_-=2.8$ msec
(these values are also given in the 7th item of Table \ref{tab:Parm}).

We note that the synaptic modification $\Delta J_{ij}^{(EI)}$ in Eq.~(\ref{eq:ItoETW})
is given by the products of Hebbian exponential functions in Eq.~(\ref{eq:Exp}) and the power function
${\Delta t_{ij}^{(EI)}}^\beta$. As in the E to E Hebbian time window, LTP occurs for $\Delta t_{ij}^{(EI)} > 0$, while LTD takes place for $\Delta t_{ij}^{(EI)} < 0$. However, due to the effect of the power function, $\Delta J_{ij}^{(EI)} \sim 0$ near
$\Delta t_{ij}^{(EI)} \sim 0$, and delayed maximum and minimum for $\Delta J_{ij}^{(EI)}$ appear at
$\Delta t_{ij}^{(EI)} = \beta \tau_+$ and $- \beta \tau_-,$ respectively. Thus, Eq.~(\ref{eq:ItoETW}) is called a time-delayed Hebbian time window, in contrast to the E to E Hebbian time window.
This time-delayed Hebbian time window was experimentally found in the case of iSTDP at inhibitory synapses (from hippocampus) onto principal excitatory stellate cells in the superficial layer II of the entorhinal cortex \cite{ItoETW1}.

For the E to I eSTDP, we employ an anti-Hebbian time window for the synaptic modification $\Delta J_{ij}^{(IE)}(\Delta t_{ij}^{(IE)})$ \cite{Abbott1,EtoITW1,EtoITW2}:
\begin{equation}
  \Delta J_{ij}(\Delta t_{ij}) = \left\{
      \begin{array}{l}
      - A_{+}~  e^{-\Delta t_{ij}^{(IE)} / \tau_{+}} ~{\rm for}~ \Delta t_{ij}^{(IE)} > 0 \\
        A_{-}~ e^{\Delta t_{ij}^{(IE)} / \tau_{-}} ~{\rm for}~ \Delta t_{ij}^{(IE)} < 0
      \end{array}
      \right. ,
\label{eq:EtoITW}
\end{equation}
where $A_+=1.0$, $A_-=0.9$, $\tau_+=15$ msec, $\tau_-=15$ msec (these values are also given in the 8th item of Table \ref{tab:Parm}), and $\Delta J_{ij}^{(IE)}(0)=0.$
For $\Delta t_{ij}^{(IE)} > 0$, LTD occurs, while LTP takes place for $\Delta t_{ij}^{(IE)} < 0$,
in contrast to the Hebbian time window for the E to E eSTDP \cite{STDP1,SSS}.
This anti-Hebbian time window was experimentally found in the case of eSTDP at excitatory synapses onto the GABAergic Purkinje-like cell in electrosensory lobe of electric fish \cite{EtoITW1}.

\subsection{Numerical Method for Integration}
\label{subsec:NM}
Numerical integration of stochastic differential Eqs.~(\ref{eq:PD1})-(\ref{eq:PD11}) with a multiplicative STDP update rule of Eqs.~(\ref{eq:MSTDP}) is done by employing the Heun method (which is developed by modifying the Euler method for the stochastic differential equations) \cite{SDE} with the time step $\Delta t=0.01$ msec. For each realization of the stochastic process, we choose random initial points $[v_i^{(X)}(0),u_i^{(X)}(0)]$ for the neuron $i$ $(i=1,\dots, N_X)$ in the $X$-population ($X=I$ or $E$) with uniform probability in the range of
$v_i^{(X)}(0) \in (-50,-45)$ and $u_i^{(X)}(0) \in (10,15)$.

\section{Effects of Interpopulation STDP on Fast Sparsely Synchronized Rhythms}
\label{sec:InterSTDP}
We consider clustered small-world networks with both I- and E-populations in Fig.~\ref{fig:CSWN}. Each Watts-Strogatz small-world network with the rewiring probability $p_{wiring}=0.25$ has high clustering and short path length due to presence of predominantly local connections and rare long-range connections.
The inhibitory small-world network consists of $N_I$ fast spiking interneurons, and the excitatory small-world network is composed of $N_E$
regular spiking pyramidal cells. Random and uniform interconnections between the inhibitory and the excitatory small-world networks are made with the
small probability $p_{inter}=1/15$. Throughout the paper, $N_I=600$ and $N_E=2400$, except for the cases in Figs.~\ref{fig:NSTDP2}(a1)-\ref{fig:NSTDP2}(a3).
Here we consider sparsely connected case. The average numbers of intrapopulation synaptic inputs per neuron
are $M_{syn}^{(I)}=40$ and $M_{syn}^{(E)}=160$, which are much smaller than $N_I$ and $N_E$, respectively.
For more details on the values of parameters, refer to Table \ref{tab:Parm}.

We first study emergence of fast sparse synchronization and its properties in the absence of STDP in the subsection \ref{subsec:NSTDP}.
Then, in the subsection \ref{subsec:ComSTDP}, we investigate the effects of interpopulation STDPs on diverse properties of population and individual behaviors of fast sparse synchronization in the combined case of both I to E iSTDP and E to I eSTDP.

\subsection{Emergence of Fast Sparse Synchronization and Its Properties in The Absence of STDP}
\label{subsec:NSTDP}
Here, we are concerned about emergence of fast sparse synchronization and its properties in the I- and the E-populations in the absence of STDP.
We also consider an interesting case of the E-I ratio balance where the ratio of average excitatory to inhibitory
synaptic strengths is the same in both fast spiking interneurons and regular spiking pyramidal cells \cite{Sparse3,Sparse4,Sparse6}.
Initial synaptic strengths are chosen from the Gaussian distribution with the mean $J_0^{(XY)}$ and the standard deviation $\sigma_0~(=5)$. The I to I synaptic strength $J_0^{(II)}(=1300)$ is strong, and hence fast sparse synchronization may appear in the I-population under the balance between strong inhibition and strong external noise. This I-population is a dominant one in our coupled two-population system because $J_0^{(II)}$ is much stronger in comparison with the E to E synaptic strength $J_0^{(EE)}(=300)$. Moreover, the I to E synaptic strength $J_0^{(EI)}=800$ is so strong that fast sparse synchronization may also appear in the E-population when the noise intensity $D$ passes a threshold. In this state of fast sparse synchronization, regular spiking pyramidal cells in the E-population make firings at much lower rates than fast spiking interneurons in the I-population. Finally, the E to I synaptic strength $J_0^{(IE)}(=487.5)$ is given by the E-I ratio balance (i.e., $J_0^{(EE)} / J_0^{(EI)}~=~J_0^{(IE)} / J_0^{(II)}$). In this subsection, all these synaptic strengths are static because we do not consider any synaptic plasticity.

\begin{figure*}
\includegraphics[width=1.7\columnwidth]{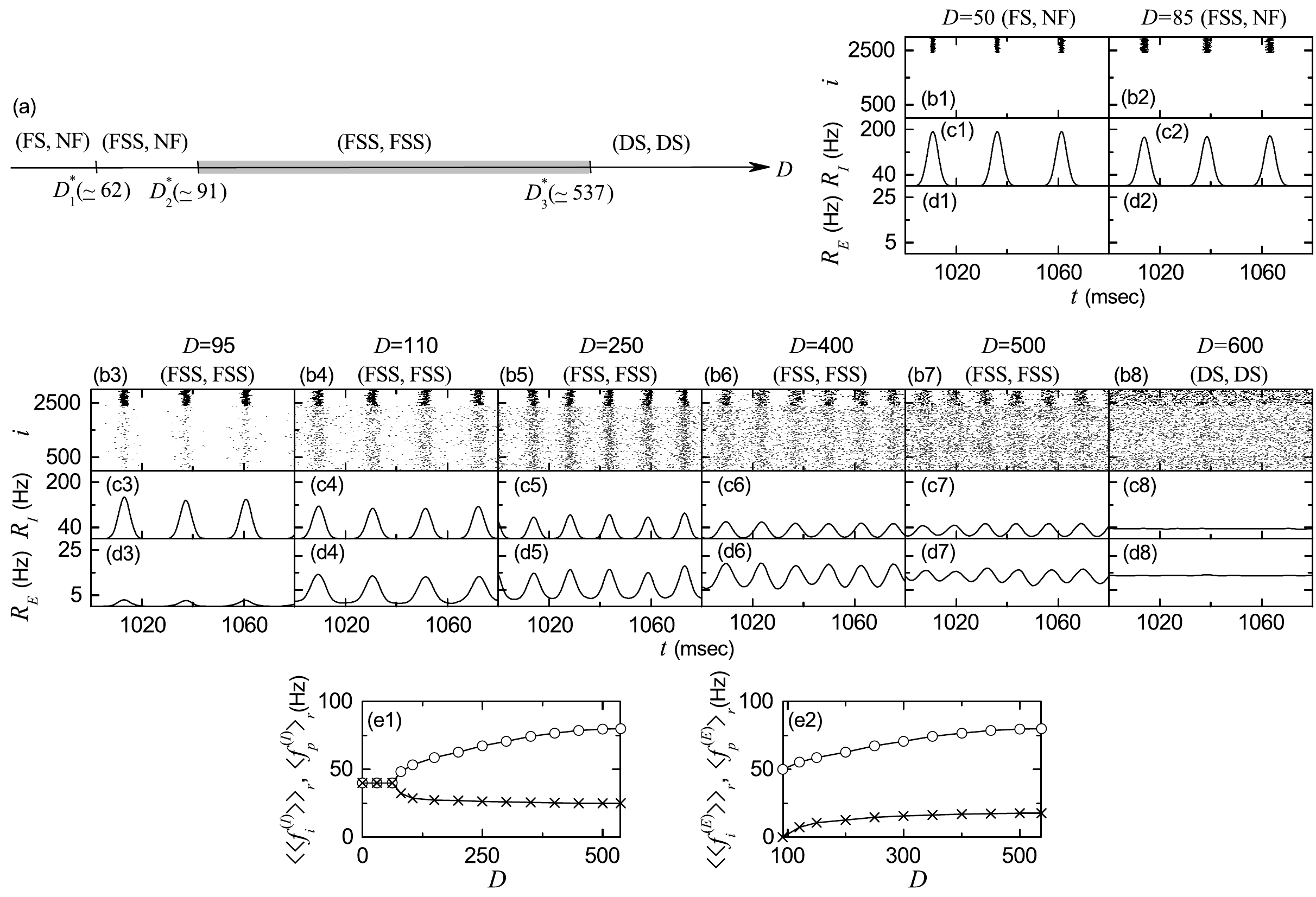}
\caption{Synchronized rhythms in both I- and E-populations in the absence of STDP. (a) Bar diagram for the population states (I, E) in the I- and E-populations.
FS, FSS, NF, and DS denote full synchronization, fast sparse synchronization, non-firing, and desynchronization, respectively.
(b1)-(b8) Raster plots of spikes for various values of $D$; lower gray dots and upper black dots denote spikes in the E- and I-populations, respectively. (c1)-(c8)
[(d1)-(d8)] Instantaneous population spike rates $R_I(t)$ [$R_E(t)$] of the I (E) population for various values of $D$. Plots of the population frequency  $\langle f_p^{(X)} \rangle_r$  (represented by open circles) and the population-averaged mean firing rates of individual neurons $\langle \langle f_i^{(X)} \rangle \rangle_r$ (denoted by crosses) versus $D$; (e1) $X=I$ (I-population) and (e2) $X=E$ (E-population).
}
\label{fig:NSTDP1}
\end{figure*}

By varying the noise intensity $D$, we investigate emergence of diverse population states in both the I- and the E-populations.
Figure \ref{fig:NSTDP1}(a) shows a bar diagram for the population states (I, E) in both I- and E-populations, where
FS, FSS, NF, and DS represents full synchronization, fast sparse synchronization, non-firing, and desynchronization, respectively.
Population synchronization may be well visualized in the raster plot of neural spikes which is a collection of spike trains of individual neurons. Such raster plots of spikes are fundamental data in experimental neuroscience. As a population quantity showing collective behaviors, we use an instantaneous population spike rate  which may be obtained from the raster plots of spikes \cite{Sparse1,Sparse2,Sparse3,Sparse4,Sparse5,Sparse6,W_Review,RM}.
For a synchronous case, ``spiking stripes'' (consisting of spikes and indicating population synchronization) are found to be formed in the raster plot, while in a desynchronized case spikes are completely scattered without forming any stripes.

Such raster plots of spikes are well shown for various values of $D$ in Figs.~\ref{fig:NSTDP1}(b1)-\ref{fig:NSTDP1}(b8).
In each raster plot, spikes of $N_I~(=600)$ fast spiking interneurons are shown with black dots in the upper part, while spikes of $N_E~(=2400)$ regular spiking  pyramidal cells are shown with gray dots in the lower part. 
Hence, in a synchronous case where spiking stripes in the raster plot appear successively at the population frequency $f_p^{(X)}$, the corresponding instantaneous population spike rate $R_X(t)$ ($X=I$ or $E$) exhibits an oscillating behavior with the population frequency $f_p^{(X)}$. On the other hand, in a desynchronized case, $R_X(t)$ is nearly stationary because spikes are completely scattered in the raster plot.
To obtain a smooth instantaneous population spike rate, we employ the kernel density estimation (kernel smoother) \cite{Kernel}. 
Each spike in the raster plot is convoluted (or blurred) with a kernel function $K_{h}(t)$ [such as a smooth Gaussian function in Eq.~(\ref{eq:Gaussian})], 
and then a smooth estimate of instantaneous population spike rate $R_{X}(t)$ is obtained by averaging the convoluted kernel function over all spikes for
all neurons in the $X$-population ($X=I$ or $E$):
\begin{equation}
R_X(t) = \frac{1}{N_X} \sum_{i=1}^{N_X} \sum_{s=1}^{n_i^{(X)}} K_h (t-t_{s}^{(i,X)}),
\label{eq:IPSR}
\end{equation}
where $t_{s}^{(i,X)}$ is the $s$th spiking time of the $i$th neuron in the $X$-population, $n_i^{(X)}$ is the total number of spikes for the $i$th neuron, and we use a Gaussian kernel function of band width $h$:
\begin{equation}
K_h (t) = \frac{1}{\sqrt{2\pi}h} e^{-t^2 / 2h^2}, ~~~~ -\infty < t < \infty.
\label{eq:Gaussian}
\end{equation}
Throughout the paper, the band width $h$ of $K_h(t)$ is 1 msec.
The instantaneous population spike rates $R_I(t)$ [$R_E(t)$] for the I-(E-)population are shown for various values of $D$ in Figs.~\ref{fig:NSTDP1}(c1)-\ref{fig:NSTDP1}(c8) [Figs.~\ref{fig:NSTDP1}(d1)-\ref{fig:NSTDP1}(d8)].

For sufficiently small $D$, individual fast spiking interneurons in the I-population fire regularly with the population-averaged mean firing rate $\langle \langle f_i^{(I)} \rangle \rangle_r$ which is the same as the population frequency $\langle f_p^{(I)} \rangle_r$ of the instantaneous population spike rate $R_I(t)$. Throughout the paper, $\langle \cdots \rangle$ denotes a population average and $\langle \cdots \rangle_r$ represents an average over 20 realizations. In this case, all fast spiking interneurons make spikings in each spiking stripe in the raster plot, and hence each stripe is fully occupied by spikes of all fast spiking interneurons. As a result, full synchronization with $\langle \langle f_i^{(I)} \rangle \rangle_r = \langle f_p^{(I)} \rangle_r$ occurs. As an example of full synchronization in the I-population, we consider the case of $D=50$. Figure \ref{fig:NSTDP1}(b1) shows the raster plot of spikes where black spiking stripes for the I-population appear successively, and the corresponding instantaneous population spike rate $R_I(t)$ with a large amplitude oscillates regularly with $\langle f_p^{(I)} \rangle_r \simeq 40$ Hz [see Fig.~\ref{fig:NSTDP1}(c1)].

In contrast, for $D=50$, regular spiking pyramidal cells in the E-population do not make firings (i.e., the E-population is in the non-firing state) due to strong I to E synaptic strength $J_0^{(EI)}$ (=800). In the isolated E-population (without synaptic coupling with the I-population), regular spiking pyramidal cells make firings with $\langle \langle f_i^{(E)} \rangle \rangle_r \simeq 189.9$ Hz in a complete incoherent way, and hence population state becomes desynchronized (i.e., in this case, spikes of regular spiking pyramidal cells are completely scattered without forming any stripes in the raster plot). However, in the presence of strong I to E synaptic current, the population state for the E-population is transformed into a non-firing state. Thus, for $D=50$ there are no spikes of regular spiking pyramidal cells in the raster plot and no instantaneous population spike rate $R_E(t)$ appears.

The full synchronization in the I-population persists until $D = D^*_1~(\simeq 62)$. For $D > D^*_1,$ full synchronization is developed into fast sparse synchronization with $\langle f_p^{(I)} \rangle_r > \langle \langle f_i^{(I)} \rangle \rangle_r$ through a pitchfork bifurcation, as shown in Fig.~\ref{fig:NSTDP1}(e1). In the case of fast sparse synchronization for $D > D^*_1,$ $\langle f_p^{(I)} \rangle_r$ ($\langle \langle f_i^{(I)} \rangle \rangle_r$) increases (decreases) monotonically from 40 Hz with increasing $D$.
In each realization, we get the population frequency $f_p^{(X)}$ ($X=I$ or $E$) from the reciprocal of the ensemble average of
$10^4$ time intervals between successive maxima of $R_X(t)$, and obtain the mean firing rate $f_i^{(X)}$ for each neuron in the $X$-population via averaging for $2 \times 10^4$ msec; $\langle f_i^{(X)} \rangle$ denotes a population-average of $f_i^{(X)}$ over all neurons in the $X$-population. Due to the noise effect, individual fast spiking interneurons fire irregularly and intermittently at lower rates than the population frequency $\langle f_p^{(I)} \rangle_r$.
Hence, only a smaller fraction of fast spiking interneurons fire in each spiking stripe in the raster plot (i.e., each spiking stripe is sparsely occupied by spikes of a smaller fraction of fast spiking interneurons).

Figures \ref{fig:NSTDP1}(b2), \ref{fig:NSTDP1}(c2), and \ref{fig:NSTDP1}(d2) show an example of fast sparse synchronization in the I-population for $D=85$. In this case, the instantaneous spike rate $R_I(t)$ of the I-population rhythm makes fast oscillations with the population frequency $\langle f_p^{(I)} \rangle_r$ ($\simeq 48.3$ Hz), while fast spiking interneurons make spikings intermittently with lower population-averaged mean firing rate $\langle \langle f_i^{(I)} \rangle \rangle_r~(\simeq$ 32.2 Hz) than the population frequency $\langle f_p^{(I)} \rangle_r$. Then, the black I-stripes (i.e., black spiking stripes for the I-population) in the raster plot become a little sparse and smeared, in comparison to the case of full synchronization for $D=50$, and hence the amplitude of the corresponding instantaneous population spike rate $R_I(t)$ (which oscillates with increased $\langle f_p^{(I)} \rangle_r$ ) also has a little decreased amplitude. Thus, fast sparsely synchronized rhythm appears in the I-population. In contrast, for $D=85$ the E-population is still in a non-firing state [see Figs.~\ref{fig:NSTDP1}(b2) and \ref{fig:NSTDP1}(d2)].

\begin{figure}
\includegraphics[width=\columnwidth]{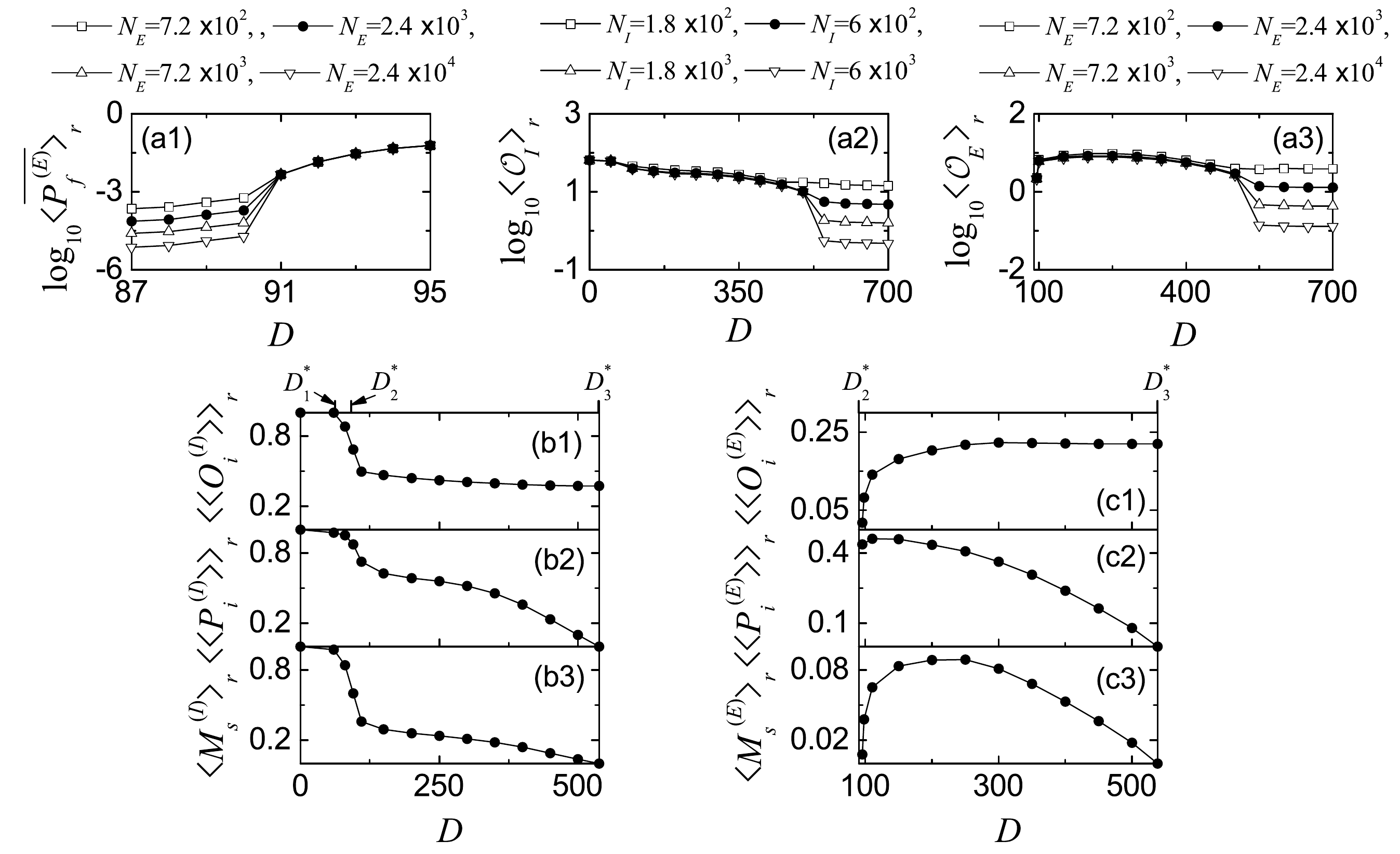}
\caption{Characterization of population synchronization in the absence of STDP. (a1) Plot of the average firing probability $\langle \overline{P_f^{(E)}} \rangle_r$ versus $D$ in the E-population. (a2) Plot of the thermodynamic order parameter $\langle {\cal{O}}_I \rangle_r$  versus $D$ in the I-population. (a3) Plot of the thermodynamic order parameter $\langle {\cal{O}}_E \rangle_r$  versus $D$ in the E-population. Plots of (b1) the average occupation degree $\langle \langle O_i^{(I)} \rangle \rangle_r$, (b2) the average pacing degree $\langle \langle P_i^{(I)} \rangle \rangle_r$, and (b3) the statistical-mechanical spiking measure $\langle M_s^{(I)} \rangle_r$  versus $D$  in the I-population. Plots of (c1) the average occupation degree $\langle \langle O_i^{(E)} \rangle \rangle_r$, (c2) the average pacing degree $\langle \langle P_i^{(E)} \rangle \rangle_r$, and (c3) the statistical-mechanical spiking measure  $\langle M_s^{(E)} \rangle_r$ versus $D$ in the E-population. In (b1)-(b3) and (c1)-(c3), $D^*_1(\simeq 62)$, $D^*_2(\simeq 91)$, and $D^*_3(\simeq 537)$ are marked on the upper horizontal axes. 
}
\label{fig:NSTDP2}
\end{figure}

However, as $D$ passes a 2nd threshold $D^*_2$ ($\simeq 91$), a transition from a non-firing to a firing state occurs in the E-population (i.e., regular spiking  pyramidal cells begin to make noise-induced intermittent spikings). [Details on this kind of firing transition will be given below in Fig.~\ref{fig:NSTDP2}(a1).] Then, fast sparse synchronization also appears in the E-population due to strong coherent I to E synaptic current to stimulate coherence between noise-induced spikings. Thus, fast sparse synchronization occurs together in both the (stimulating) I- and the (stimulated) E-populations, as shown in the raster plot of spikes in Fig.~\ref{fig:NSTDP1}(b3) for $D=95$. The instantaneous population spike rates $R_I(t)$ and $R_E(t)$ for the sparsely synchronized rhythms in the I- and the E-populations oscillate fast with the same population frequency $\langle f_p^{(I)} \rangle_r~ = ~ \langle f_p^{(E)} \rangle_r~ (\simeq$ 51.3 Hz).
Here, we note that the population frequency of fast sparsely synchronized rhythms is determined by the dominant stimulating I-population, and hence $\langle f_p^{(E)} \rangle_r$ for the E-population is just the same as $\langle f_p^{(I)} \rangle_r$ for the I-population. However, regular spiking pyramidal cells fire intermittent spikings with much lower population-averaged mean firing rate $\langle \langle f_i^{(E)} \rangle \rangle_r$ ($\simeq 2.7$ Hz) than $\langle \langle f_i^{(I)} \rangle \rangle_r$ ($\simeq 30$ Hz) of fast spiking interneurons. Hence, the gray E-stripes (i.e., gray spiking stripes for the E-population) in the raster plot of spikes are much more sparse than the black I-stripes, and the amplitudes of $R_E(t)$ are much smaller than those of $R_I(t)$.

With further increasing $D$, we study evolutions of (FSS, FSS) in both the I- and the E-populations for various values of
$D$ ($D=$110, 250, 400, and 500). For these cases, raster plots of spikes are shown in Figs.~\ref{fig:NSTDP1}(b4)-\ref{fig:NSTDP1}(b7), and
instantaneous population spike rates $R_I(t)$ and $R_E(t)$ are given in Figs.~\ref{fig:NSTDP1}(c4)-\ref{fig:NSTDP1}(c7) and Figs.~\ref{fig:NSTDP1}(d4)-\ref{fig:NSTDP1}(d7), respectively.
In the I-population, as $D$ is increased, more number of black I-stripes appear successively in the raster plots, which implies increase in the population frequency $\langle f_p^{(I)} \rangle_r$ [see Fig.~\ref{fig:NSTDP1}(e1)].
Furthermore, these black I-stripes become more sparse (i.e., density of spikes in the black I-stripes decreases) due to decrease in $\langle \langle f_i^{(I)} \rangle \rangle_r$ [see Fig.~\ref{fig:NSTDP1}(e1)], and they also are more and more smeared. Hence, with increasing $D$ monotonic decrease in amplitudes of the corresponding instantaneous population spike rate $R_I(t)$ occurs (i.e. the degree of fast sparse synchronization in the I-population is decreased). Eventually, when passing the 3rd  threshold $D^*_3$ ($\simeq$ 537), a transition from fast sparse synchronization to desynchronization occurs because of complete overlap between black I-stripes in the raster plot. Then, spikes of fast spiking interneurons are completely scattered in the raster plot, and the instantaneous population spike rate $R_I(t)$ is nearly stationary, as shown in Figs.~\ref{fig:NSTDP1}(b8) and \ref{fig:NSTDP1}(c8) for $D=600$.

In the E-population, the instantaneous population spike rate $R_E(t)$ for the sparsely synchronized rhythm oscillates fast with the population frequency
$\langle f_p^{(E)} \rangle_r$ which is the same as $\langle f_p^{(I)} \rangle_r$ for the I-population; $\langle f_p^{(E)} \rangle_r$ increases with $D$ [see Fig.~\ref{fig:NSTDP1}(e2)]. As $D$ is increased, population-averaged mean firing rate $\langle \langle f_i^{(E)} \rangle \rangle_r$ also increases due to decrease in the coherent I to E synaptic current (which results from decrease in the degree of fast sparse synchronization in the I-population) [see Fig.~\ref{fig:NSTDP1}(e2)], in contrast to the case of $\langle \langle f_i^{(I)} \rangle \rangle_r$ in the I-population (which decreases with $D$). Hence, as $D$ is increased, density of spikes in gray E-stripes in the raster plot increases (i.e., gray E-stripes become less sparse), unlike the case of I-population. On the other hand, with increasing $D$ for $D>110$ E-stripes are more and more smeared, as in the case of I-population.

The degree of fast sparse synchronization is determined by considering both the density of spikes [denoting the average occupation degree (corresponding to average fraction of regular spiking pyramidal cells in each E-stripe)] and the pacing degree of spikes (representing the degree of phase coherence between spikes) in the E-stripes, the details of which will be given in Fig.~\ref{fig:NSTDP2}. Through competition between the (increasing) occupation degree and the (decreasing) pacing degree, it is found that the E-population has the maximum degree of fast sparse synchronization for $D \sim 250$; details on the degree of fast sparse synchronization will be given below in Fig.~\ref{fig:NSTDP2}. Thus, the amplitude of $R_E(t)$ (representing the overall degree of fast sparse synchronization) increases until $D \sim 250$, and then it decreases monotonically. Like the case of I-population, due to complete overlap between the gray E-stripes in the raster plot, a transition to desynchronization occurs at the same 3rd threshold $D^*_3$. Then, spikes of
regular spiking pyramidal cells are completely scattered in the raster plot and the instantaneous population spike rate $R_E(t)$ is nearly stationary [see Figs.~\ref{fig:NSTDP1}(b8) and \ref{fig:NSTDP1}(d8) for $D=600$].

For characterization of fast sparse synchronization [shown in Figs.~\ref{fig:NSTDP1}(b3)-\ref{fig:NSTDP1}(b8)], we first determine the
2nd and 3rd thresholds $D^*_2~(\simeq 91)$ and $D^*_3~(\simeq 537)$. When passing the 2nd threshold $D^*_2$, a firing transition
(i.e., transition from a non-firing to a firing state) occurs in the E-population. We quantitatively characterize
this firing transition in terms of the average firing probability $\overline{P_f^{(E)}}$ \cite{AFP}.
In each raster plot of spikes in the E-population, we divide a long-time interval into bins of width $\delta~(=5$ msec) and
calculate the firing probability in each $i$th bin (i.e., the fraction of firing regular spiking pyramidal cells in the $i$th bin):
\begin{equation}
P_f^{(E)}(i) = \frac {N_f^{(E)}(i)} {N_E},~~ i=1,~2,~ \cdots,
\label{eq:PF}
\end{equation}
where $N_f^{(E)}(i)$ is the number of firing regular spiking pyramidal cells in the $i$th bin.
Then, we get the average firing probability $\overline{P_f^{(E)}}$ via time average of
$P_f^{(E)}(i)$ over sufficiently many bins:
\begin{equation}
\overline{P_f^{(E)}} = {\frac {1} {N_b}} \sum_{i=1}^{N_b} P_f^{(E)}(i),
\label{eq:AFP}
\end{equation}
where $N_b$ is the number of bins for averaging.
In each realization, the averaging is done for sufficiently large number of bins ($N_b=4000$).
For a firing (non-firing) state, the average firing probability $\overline{P_f^{(E)}}$ approaches
a non-zero (zero) limit value in the thermodynamic limit of $N_E \rightarrow \infty$.

Figure \ref{fig:NSTDP2}(a1) shows a plot of $\log_{10} \langle \overline{P_f^{(E)}} \rangle_r$ versus the noise intensity $D$.
For $D > D^*_2~(\simeq 91)$, firing states appear in the E-population (i.e., regular spiking pyramidal cells make noise-induced intermittent spikings) because $\langle \overline{P_f^{(E)}} \rangle_r$ tends to converge toward non-zero limit values.
Then, strong coherent I to E synaptic input current stimulates fast sparse synchronization between these noise-induced intermittent spikes
in the E-population. Thus, when passing the 2nd threshold $D_2^*,$ (FSS, FSS) occurs in both the I- and the E-populations.

However, as $D$ is further increased, the degree of (FSS, FSS) decreases, and eventually when passing the 3rd threshold $D^*_3~(\simeq 537)$,
a transition to desynchronization occurs in both the I- and the E-populations, due to a destructive role of noise to spoil fast sparse synchronization.
We characterize this kind of synchronization-desynchronization transition in the $X$-population ($X=I$ or $E$) in terms of the order parameter ${\cal {O}}_X,$ corresponding to the mean square deviation of the instantaneous population spike rate $R_X(t)$ \cite{RM}:
\begin{equation}
{\cal{O}}_X = \overline{ (R_X(t) - \overline{R_X(t)})^2 }.
\label{eq:Order}
\end{equation}
This order parameter may be regarded as a thermodynamic measure because it concerns just the macroscopic instantaneous population spike rate $R_X(t)$
without any consideration between $R_X(t)$ and microscopic individual spikes.
For a synchronized state, $R_X(t)$ exhibits an oscillatory behavior, while for a desynchronized state it is nearly stationary.
Hence, the order parameter ${\cal{O}}_X$ approaches a non-zero (zero) limit value in the synchronized (desynchronized) case in the thermodynamic limit of $N_X \rightarrow \infty$.
In each realization, we obtain ${\cal{O}}_X$ by following a stochastic trajectory for $3 \times 10^4$ msec.

Figures \ref{fig:NSTDP2}(a2) and \ref{fig:NSTDP2}(a3) show plots of $\log_{10} \langle {\cal O}_I \rangle_r$ and
$\log_{10} \langle {\cal O}_E \rangle_r$ versus $D$, respectively.
For $D < D^*_3$ ($\simeq 537$), (FSS, FSS) occurs in both the I- and the E-populations because the order parameters $\langle {\cal{O}}_I \rangle_r$ and $\langle {\cal{O}}_E \rangle_r$ tend to converge toward non-zero limit values.
In contrast, for $D > D^*_3$, with increasing $N_I$ and $N_E$ both the order parameters $\langle {\cal{O}}_I \rangle_r$ and
$\langle {\cal{O}}_E \rangle_r$ tend to approach zero, and hence a transition to desynchronization occurs together in both the I- and the E-populations.

We now measure the degree of fast sparse synchronization in the I- and the E-populations by employing the statistical-mechanical spiking measure $M_s^{(X)}$
($X=I$ or $E$) \cite{RM}. This spiking measure $M_s^{(X)}$ has been successively applied
for characterization of various types of spike and burst synchronizations \cite{FSS-SWN,FSS-SFN,FSS-CSWN,SSS,SBS,BS-iSTDP,FSS-iSTDP,RM,BS-SWN,BS-MS,BS-SFN,IHSWN,EI-SS}.
For a synchronous case, spiking I-(E-)stripes appear successively in the raster plot of spikes of fast spiking interneurons (regular spiking pyramidal cells). The spiking measure $M_i^{(X)}$ of the $i$th $X-$stripe is defined by the product of the occupation degree $O_i^{(X)}$ of spikes
(denoting the density of the $i$th $X-$stripe) and the pacing degree $P_i^{(X)}$ of spikes (representing the degree of phase coherence between spikes in the $i$th $X-$stripe):
\begin{equation}
M_i^{(X)} = O_i^{(X)} \cdot P_i^{(X)}.
\label{eq:SMi}
\end{equation}
The occupation degree $O_i^{(X)}$ of spikes in the $X-$stripe is given by the fraction of spiking neurons:
\begin{equation}
   O_i^{(X)} = \frac {N_i^{(s,X)}} {N_X},
\label{eq:OCCU}
\end{equation}
where $N_i^{(s,X)}$ is the number of spiking neurons in the $i$th $X-$stripe.
In the case of sparse synchronization, $O_i^{(X)}<1$, in contrast to the case of full synchronization with $O_i^{(X)}=1$.

The pacing degree $P_i^{(X)}$ of spikes in the $i$th $X-$stripe can be determined in a statistical-mechanical way by considering their contributions to the macroscopic instantaneous population spike rate $R_X(t)$.
Central maxima of $R_X(t)$ between neighboring left and right minima of $R_X(t)$ coincide with centers of $X-$stripes in the raster plot. A global cycle begins from a left minimum of $R_X(t)$, passes a maximum, and ends at a right minimum. An instantaneous global phase $\Phi^{(X)}(t)$ of $R_X(t)$ was introduced via linear interpolation in the region forming a global cycle [for details, refer to Eqs.~(16) and (17) in \cite{RM}]. Then, the contribution of the $k$th microscopic spike in the $i$th $X-$stripe occurring at the time $t_k^{(s,X)}$ to $R_X(t)$ is
given by $\cos \Phi_k^{(X)}$, where $\Phi_k^{(X)}$ is the global phase at the $k$th spiking time [i.e., $\Phi_k^{(X)} \equiv \Phi^{(X)}(t_k^{(s,X)})$]. A microscopic spike makes the most constructive (in-phase)
contribution to $R_X(t)$ when the corresponding global phase $\Phi_k^{(X)}$ is $2 \pi n$ ($n=0,1,2, \dots$). In contrast, it makes the most destructive (anti-phase) contribution to $R_X(t)$ when $\Phi_k^{(X)}$
is $2 \pi (n-1/2)$. By averaging the contributions of all microscopic spikes in the $i$th $X-$stripe to $R_X(t)$, we get the pacing degree of spikes in the $i$th $X-$stripe [refer to Eq.~(18) in \cite{RM}]:
\begin{equation}
 P_i^{(X)} ={ \frac {1} {S_i^{(X)}}} \sum_{k=1}^{S_i^{(X)}} \cos \Phi_k^{(X)},
\label{eq:PACING}
\end{equation}
where $S_i^{(X)}$ is the total number of microscopic spikes in the $i$th $X-$stripe.
Then, via averaging $M_i^{(X)}$ of Eq.~(\ref{eq:SMi}) over a sufficiently large number $N_s^{(X)}$ of $X-$stripes, we
obtain the statistical-mechanical spiking measure $M_s^{(X)}$, based on the instantaneous population spike rate $R_X(t)$
[refer to Eq.~(19) in \cite{RM}]:
\begin{equation}
M_s^{(X)} =  {\frac {1} {N_s^{(X)}}} \sum_{i=1}^{N_s^{(X)}} M_i^{(X)}.
\label{eq:SM}
\end{equation}
In each realization, we obtain $\langle O_i^{(X)} \rangle,$ $\langle P_i^{(X)} \rangle,$ and $M_s^{(X)}$ by following
$6 \times 10^3$ $X-$stripes.

We first consider the case of I-population (i.e., $X= I$) which is a dominant one in our
coupled two-population network.
Figures \ref{fig:NSTDP2}(b1)-\ref{fig:NSTDP2}(b3) show the average occupation degree $\langle \langle O_i^{(I)} \rangle
\rangle_r$, the average pacing degree $\langle \langle P_i^{(I)} \rangle \rangle_r$, and the statistical-mechanical spiking
measure $\langle M_s^{(I)} \rangle_r$ in the range of $0 < D < D^*_3$, respectively.
With increasing $D$ from 0 to $D_1^*~(\simeq 62)$, full synchronization persists, and hence $\langle \langle O_i^{(I)} \rangle \rangle_r =1$. In this range of $D$, $\langle \langle P_i^{(I)} \rangle \rangle_r$ decreases very slowly from 1.0 to 0.98.
In the case of full synchronization, the statistical-mechanical spiking measure is equal to the average pacing degree (i.e., $\langle M_s^{(I)} \rangle_r = \langle \langle P_i^{(I)} \rangle \rangle_r$).
However, as $D$ is increased from $D^*_1$, full synchronization is developed into fast sparse synchronization. In the case of fast sparse synchronization, at first $\langle \langle O_i^{(I)} \rangle \rangle_r$ (representing the density of spikes in the I-stripes) decreases rapidly due to break-up of full synchronization, and then it slowly decreases toward a limit value of $\langle \langle O_i^{(I)} \rangle \rangle_r \simeq 0.37$ for $D=D^*_3$, like the behavior of population-averaged mean firing rate $\langle \langle f_i^{(I)} \rangle \rangle_r$ in Fig.~\ref{fig:NSTDP1}(e1). The average pacing degree $\langle \langle P_i^{(I)} \rangle \rangle_r$ denotes well the
average degree of phase coherence between spikes in the I-stripes; as the I-stripes become more smeared, their pacing degree gets decreased. With increasing $D$, $\langle \langle P_i^{(I)} \rangle \rangle_r$ decreases due to intensified smearing,
and for large $D$ near $D^*_3$ it converges to zero due to complete overlap between sparse spiking I-stripes.
The statistical-mechanical spiking measure $\langle M_s^{(I)} \rangle_r$ is obtained via product of the occupation and the pacing degrees of spikes. Due to the rapid decrease in $\langle \langle O_i^{(I)} \rangle \rangle_r$, at first $\langle M_s^{(I)} \rangle_r$ also decreases rapidly, and then it makes a slow convergence to zero for $D=D^*_3$, like the case of
$\langle \langle P_i^{(I)} \rangle \rangle_r$. Thus, three kinds of downhill-shaped curves (composed of solid circles) for $\langle \langle O_i^{(I)} \rangle \rangle_r$, $\langle \langle P_i^{(I)} \rangle \rangle_r$ and $\langle M_s^{(I)} \rangle_r$ are formed [see Figs.~\ref{fig:NSTDP2}(b1)-\ref{fig:NSTDP2}(b3)].

Figures \ref{fig:NSTDP2}(c1)-\ref{fig:NSTDP2}(c3) show $\langle \langle O_i^{(E)} \rangle \rangle_r$, $\langle \langle P_i^{(E)} \rangle \rangle_r$, and $\langle M_s^{(E)} \rangle_r$ in the E-population for $D^*_2 < D < D^*_3$, respectively.
When passing the 2nd threshold $D^*_2$, fast sparse synchronization appears in the E-population because strong coherent I to E synaptic input current stimulates coherence between noise-induced intermittent spikes [i.e., sparsely synchronized E-population rhythms are locked to (stimulating) sparsely synchronized I-population rhythms]. In this case, at first, the average occupation degree $\langle \langle O_i^{(E)} \rangle \rangle_r$ begins to make a rapid increase from 0, and then it increases slowly to a saturated
limit value of $\langle \langle O_i^{(E)} \rangle \rangle_r \simeq 0.22$. Thus, an uphill-shaped curve for $\langle \langle O_i^{(E)} \rangle \rangle_r$ is formed, similar to the case of population-averaged mean firing rate $\langle \langle f_i^{(E)} \rangle \rangle_r$ in Fig.~\ref{fig:NSTDP1}(e2).
In contrast, just after $D=D^*_2$, the average pacing degree $\langle \langle P_i^{(E)} \rangle \rangle_r$ starts from a non-zero value (e.g., $\langle \langle P_i^{(E)} \rangle \rangle_r \simeq 0.409$ for $D=92$), it increases to a maximum value ($\simeq 0.465$) for $D \sim 150$, and then it decreases monotonically to zero at the 3rd threshold $D^*_3$ because of
complete overlap between sparse E-stripes. Thus, for $D > 150$ the graph for $\langle \langle P_i^{(E)} \rangle \rangle_r$ is a downhill-shaped curve. Through the product of the occupation (uphill curve) and the pacing (downhill curve) degrees, the spiking measure $\langle M_s^{(E)} \rangle_r$ forms a bell-shaped curve with a maximum ($\simeq 0.089$) at $D \sim 250$; the values of $\langle M_s^{(E)} \rangle_r$ are zero at both ends ($D^*_2$ and $D^*_3$). This spiking measure
$\langle M_s^{(E)} \rangle_r$ of the E-population rhythms is much less than that $\langle M_s^{(I)} \rangle_r$ of the
dominant I-population rhythms.

\begin{figure}
\includegraphics[width=\columnwidth]{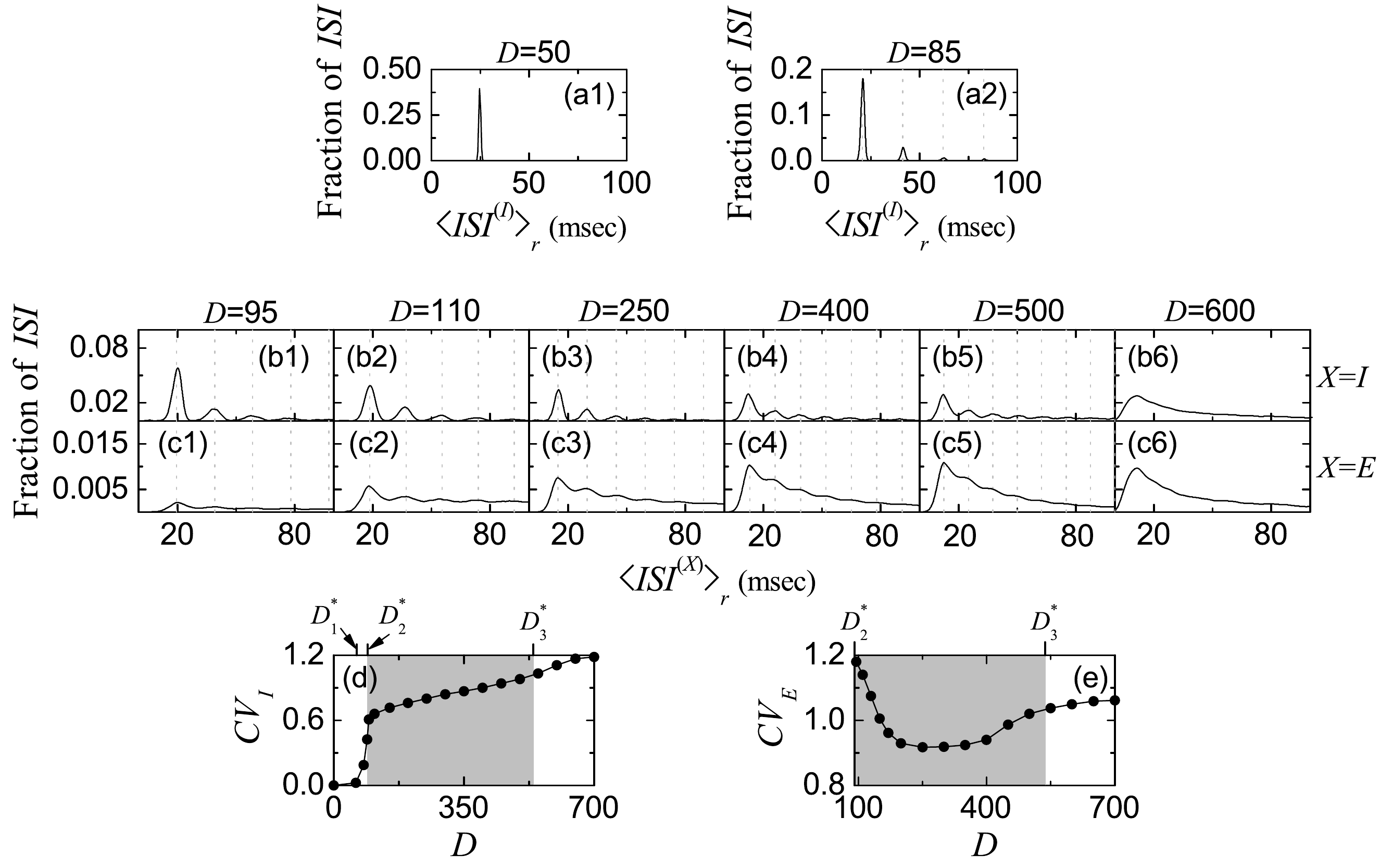}
\caption{Characterization of individual spiking behaviors in the absence of STDP. ISI histograms for $D=$ (a1) 50 and (a2) 85 in the I-population. ISI histograms for various values of $D$ in the I-population ($X=I$) [(b1)-(b6)] and the E-population ($X=E$) [(c1)-(c6)].
Vertical dotted lines in (a2), (b1)-(b5), and (c1)-(c5) represent multiples of the global period $T_G^{(X)}$ of the instantaneous population spike rate
$R_X(t)$ ($X=I$ or $E$). Plots of the coefficient of variation $CV_X$ versus $D$; $X=$ (d) $I$ and (e) $E$.
In (d) and (e), (FSS, FSS) occurs in the intermediate gray-shaded region, and $D^*_1(\simeq 62)$, $D^*_2(\simeq 91)$, and $D^*_3(\simeq 537)$ are marked 
on the upper horizontal axes.
}
\label{fig:NSTDP3}
\end{figure}

In addition to characterization of population synchronization in Fig.~\ref{fig:NSTDP2}, we also characterize individual spiking behaviors of
fast spiking interneurons and regular spiking pyramidal cells in terms of interspike intervals (ISIs) in Fig.~\ref{fig:NSTDP3}. In each realization, we obtain one
ISI histogram which is composed of $10^5$ ISIs obtained from all individual neurons, and then we get an averaged
ISI histogram for $\langle ISI^{(X)} \rangle_r$ ($X=$ $I$ or $E$) via 20 realizations.

We first consider the case of (stimulating) dominant I-population. In the case of full synchronization for $D=50$, the ISI  histogram is shown in Fig.~\ref{fig:NSTDP3}(a1). It has a sharp single peak at $\langle ISI^{(I)} \rangle_r \simeq 25$ msec. In this case, all fast spiking interneurons exhibit regular spikings like clocks with $\langle \langle  f_i^{(I)} \rangle \rangle_r \simeq 40$ Hz, which leads to emergence of fully synchronized rhythm with the same population frequency $\langle f_p^{(I)} \rangle \simeq 40$ Hz.

However, when passing the 1st threshold $D^*_1~(\simeq 62),$ fast sparse synchronization emerges via break-up of full synchronization due to a destructive role of noise. Due to the noise effect, individual fast spiking interneurons exhibit intermittent spikings phase-locked to the instantaneous population spike rate $R_I(t)$ at random multiples of the global period $T_G^{(I)}$ of $R_I(t)$, unlike the case of full synchronization. This ``stochastic phase locking,'' resulting in ``stochastic spike skipping,'' is well shown in the ISI histogram with multiple peaks appearing at integer multiples of $T_G^{(I)}$, as shown in Fig.~\ref{fig:NSTDP3}(a2) for $D=85$, which is in contrast to the case of full synchronization with a single-peaked ISI histogram. In this case, the 1st-order main peak at $T_G^{(I)}$ ($\simeq 20.7$ msec) is a dominant one, and smaller 2nd- and 3rd-order peaks (appearing at $2~T_G^{(I)}$ and $3~T_G^{(I)}$) may also be seen. Here, vertical dotted lines in Fig.~\ref{fig:NSTDP3}(a2), Figs.~\ref{fig:NSTDP3}(b1)-\ref{fig:NSTDP3}(b5), and
Figs.~\ref{fig:NSTDP3}(c1)-\ref{fig:NSTDP3}(c5) represent multiples of the global period $T_G^{(X)}$ of the instantaneous population spike rate $R_X(t)$
($X=$ $I$ or $E$). In the case of $D=85$, the average ISI $\langle \langle ISI^{(I)} \rangle_r \rangle$ ($\simeq 31.0$ msec) is increased, in comparison with that in the case of full synchronization. Hence, fast spiking interneurons make intermittent spikings at lower population-averaged mean firing rate $\langle \langle f_i^{(I)} \rangle_r \rangle~(\simeq 32.3$ Hz) than the population frequency $\langle f_p^{(I)} \rangle_r$ ($\simeq 48.3$ Hz), in contrast to the case of full synchronization with $\langle \langle f_i^{(I)} \rangle \rangle_r = \langle f_p^{(I)} \rangle_r$ ($\simeq$ 40 Hz).

This kind of spike-skipping phenomena (characterized with multi-peaked ISI histograms) have also been found in networks of coupled inhibitory neurons where noise-induced hoppings from one cluster to another one occur \cite{GR}, in single noisy neuron models exhibiting stochastic resonance due to a weak periodic external force \cite{Longtin1,Longtin2}, and in inhibitory networks of coupled subthreshold neurons showing stochastic spiking coherence \cite{Kim1,Kim2,Kim3}.
Because of this stochastic spike skipping, the population-averaged mean firing rate of individual neurons becomes less than the population frequency, which leads to occurrence of sparse synchronization (i.e., sparse occupation occurs in spiking stripes in the raster plot).

As $D$ passes the 2nd threshold $D^*_2$ ($\simeq 91$), fast sparse synchronization emerges in the E-population because of strong coherent I to E synaptic input current stimulating coherence between noise-induced intermittent spikes. Thus, for $D > D^*_2$ fast sparse synchronization occurs together in both the I- and the E-populations. However, when passing the large 3rd threshold $D^*_3$ ($\simeq 537$), a transition from fast sparse synchronization to desynchronization occurs due to a destructive role of noise to spoil fast sparse synchronization. Hence, for $D>D^*_3$ desynchronized states exist in both the I- and the E-populations. With increasing $D$ from $D^*_2$, we investigate individual spiking behaviors in terms of ISIs in both the I- and the E-populations.

Figures \ref{fig:NSTDP3}(b1)-\ref{fig:NSTDP3}(b5) show ISI histograms for various values of $D$ in the case of fast sparse synchronization in the (stimulating) dominant I-population. Due to the stochastic spike skippings, multiple peaks appear at integer multiples of the global period $T_G^{(I)}$ of $R_I(t)$.
As $D$ is increased, fast spiking interneurons tend to fire more irregularly and sparsely.
Hence, the 1st-order main peak becomes lowered and broadened, higher-order peaks also become wider, and thus mergings between multiple peaks occur. Hence, with increasing $D$, the average ISI $\langle \langle ISI^{(I)} \rangle_r \rangle$ increases due to developed tail part.
We note that the population-averaged mean firing rate $\langle \langle f_i^{(I)} \rangle \rangle_r$ corresponds to the reciprocal of the  average ISI $\langle \langle ISI^{(I)} \rangle_r \rangle$. Hence, as $D$ is increased in the case of fast sparse synchronization,
$\langle \langle f_i^{(I)} \rangle \rangle_r$ decreases [see Fig.~\ref{fig:NSTDP1}(e1)].
These individual spiking behaviors make some effects on population behaviors.
Because of decrease in $\langle \langle f_i^{(I)} \rangle \rangle_r$ with increasing $D$, spikes become more sparse, and hence the average occupation degree
$\langle \langle O_i^{(I)} \rangle \rangle_r$ in the spiking stripes in the raster plots decreases, as shown in Figs.~\ref{fig:NSTDP2}(b1).
Also, due to merging between peaks (i.e., due to increase in the irregularity degree of individual spikings), spiking stripes in the raster plots in Figs.~\ref{fig:NSTDP1}(b3)-\ref{fig:NSTDP1}(b7) become more smeared as $D$ is increased, and hence the average pacing degrees $\langle \langle P_i^{(I)} \rangle \rangle_r$ of spikes in the stripes get decreased [see Fig.~\ref{fig:NSTDP2}(b2)].

Eventually, when passing the 3rd threshold $D^*_3 (\simeq 537)$, a desynchronized state where spikes in the raster plot are completely scattered. In this case of desynchronization, a broad single peak appears in the ISI histogram due to complete overlap of multiple peaks. Thus, for $D=600$ a single-peaked ISI histogram with a long tail appears, as shown in Fig.~\ref{fig:NSTDP3}(b6). In this case of $D=600$, the average ISI $\langle \langle ISI^{(I)} \rangle_r \rangle$ ($\simeq 39.7$ msec) is a little shorter than that ($\simeq 40$ msec) for $D=500$, in contrast to the increasing tendency in the case of fast sparse synchronization.
In the desynchronized state for $D > D^*_3,$ the I to I synaptic current is incoherent (i.e., the instantaneous population spike rate $R_I(t)$ is nearly stationary), and hence noise no longer causes stochastic phase lockings. In this case, noise just makes fast spiking interneurons fire more frequently, along with the incoherent synaptic input currents. Thus, with increasing $D$ in the desynchronized case, the average ISI $\langle \langle ISI^{(I)} \rangle_r \rangle$ tends to decrease, in contrast to the case of fast sparse synchronization.
The corresponding population-averaged mean firing rate $\langle \langle f_i^{(I)} \rangle \rangle_r$ in the desynchronized case also tends to increase, in contrast to the decreasing tendency in the case of fast sparse synchronization.

We now consider the case of (stimulated) E-population for $D>D^*_2$.
Figures \ref{fig:NSTDP3}(c1)-\ref{fig:NSTDP3}(c5) show ISI histograms for various values of $D$ in the
case of fast sparse synchronization.
In this case, both the coherent I to E synaptic input and noise make effects on individual spiking behaviors of regular spiking pyramidal cells.
Due to the stochastic spike skippings, multiple peaks appear, as in the case of I-population.
However, as $D$ is increased, stochastic spike skippings become weakened (i.e., regular spiking pyramidal cells tend to fire less sparsely)
due to decrease in strengths of the stimulating I to E synaptic input currents.
Hence, the heights of major lower-order peaks (e.g. the main 1st-order peak and the 2nd- and 3rd-order peaks) continue to increase with increasing $D$, in contrast to the case of I-population where the major peaks are lowered due to noise effect.

Just after appearance of fast sparse synchronization (appearing due to coherent I to E synaptic current), a long tail is developed so much in the ISI histogram [e.g., see Fig.~\ref{fig:NSTDP3}(c1) for $D=95$], and hence multiple peaks are less developed. As $D$ is a little more increased, multiple peaks begin to be clearly developed due to a constructive role of coherent I to E synaptic input, as shown in Fig.~\ref{fig:NSTDP3}(c2) for $D=110$.
Thus, the average pacing degree $\langle \langle P_i^{(E)} \rangle \rangle_r$ of spikes in the E-stripes for $D=110$ increases a little in comparison with that for $D=95$, as shown in Fig.~\ref{fig:NSTDP2}(c2). However, as $D$ is further increased for $D>150$, mergings between multiple peaks begin to occur due to a destructive role of noise, as shown in Figs.~\ref{fig:NSTDP3}(c3)-\ref{fig:NSTDP3}(c5). Hence, with increasing $D$ from $150$, the average pacing degree $\langle \langle P_i^{(E)} \rangle \rangle_r$ of spikes also begins to decrease [see Fig.~\ref{fig:NSTDP2}(c2)], as in the case of I-population.

With increasing $D$ in the case of fast sparse synchronization, the average ISI $\langle \langle ISI^{(E)} \rangle_r \rangle$ decreases mainly due to increase in the heights of major lower-order peaks, in contrast to the increasing tendency for $\langle \langle ISI^{(I)} \rangle_r \rangle$ in the I-population. This decreasing tendency for $\langle \langle ISI^{(E)} \rangle_r \rangle$ continues even in the case of desynchronization. Figure \ref{fig:NSTDP3}(c6) shows a single-peaked ISI histogram with a long tail (that appears through complete merging between multiple peaks) for $D=600$ (where desynchronization occurs). In this case, the average ISI $\langle \langle ISI^{(E)} \rangle_r \rangle$ ($\simeq 54.9$ msec) is shorter than that ($ 56.8$ msec) in the case of fast sparse synchronization for $D=500$. We also note that for each value of $D$ (in the case of fast sparse synchronization and desynchronization), $\langle \langle ISI^{(E)} \rangle_r \rangle$ is longer than $\langle \langle ISI^{(I)} \rangle_r \rangle$ in the case of I-population, due to much more developed tail part.

As a result of decrease in the average ISI $\langle \langle ISI^{(E)} \rangle_r \rangle$, the population-averaged mean firing rate $\langle \langle f_i^{(E)} \rangle \rangle_r$ (corresponding to the reciprocal of $\langle \langle ISI^{(E)} \rangle_r \rangle$) increases with $D$ [see Fig.~\ref{fig:NSTDP1}(e2)]. We also note that these population-averaged mean firing rates $\langle \langle f_i^{(E)} \rangle \rangle_r$ are much lower than $\langle \langle f_i^{(I)} \rangle \rangle_r$ in the (stimulating) I-population, although the population frequencies in both populations are the same. In the case of fast sparse synchronization, due to increase in $\langle \langle f_i^{(E)} \rangle \rangle_r$ with $D$, E-stripes in the raster plot become less sparse [i.e., the average occupation degree $\langle \langle O_i^{(E)} \rangle \rangle_r$ of spikes in the E-stripes increases, as shown in Fig.~\ref{fig:NSTDP2}(c1)].
The increasing tendency for $\langle \langle f_i^{(E)} \rangle \rangle_r$ continues even in the case of desynchronization.
For example, the population-averaged mean firing rate $\langle \langle f_i^{(E)} \rangle \rangle_r~(\simeq 18.2$ Hz) for $D=600$ is increased in comparison with that ($\simeq 17.6$ Hz) for $D=500$.

We are also concerned about temporal variability of individual single-cell discharges. Such temporal variability of individual single-cell firings may be characterized in terms of the coefficient of variation for the distribution of ISIs (defined by the ratio of the standard deviation to the mean for
the ISI distribution) \cite{CV1}. The larger the coefficient of variation is, the more irregular single-cell firings get. Using this coefficient of variation, spiking and bursting patterns have been well characterized \cite{CV2}.
As the coefficient of variation is increased, the irregularity degree of individual firings of single cells increases. For example, in the case of a Poisson process, the coefficient of variation takes a value 1. However, this (i.e., to take the value 1 for the coefficient of variation) is just a necessary, though not sufficient, condition to identify a Poisson spike train. When the coefficient of variation for a spike train is less than 1, it is more regular than a Poisson process with the same mean firing rate \cite{CV3}. On the other hand, if the coefficient of variation is larger than 1, then the spike train is more irregular than the Poisson process (e.g., see Fig.~1C in \cite{Sparse1}). By varying $D$, we obtain coefficients of variation from the realization-averaged ISI histograms.
Figures \ref{fig:NSTDP3}(d) and \ref{fig:NSTDP3}(e) show plots of the coefficients of variation, $CV_I$ and $CV_E$, versus $D$ for individual firings of fast spiking interneurons (I-population) and regular spiking pyramidal cells (E-population), respectively. Gray-shaded regions in Figs.~\ref{fig:NSTDP3}(d)-\ref{fig:NSTDP3}(e) correspond to the regions of (FSS, FSS) (i.e., $D^*_2 < D < D^*_3$) where fast sparse synchronization appears in both the I- and the E-populations.

We first consider the case of fast spiking interneurons in Fig.~\ref{fig:NSTDP3}(d). In the case of full synchronization for $D < D^*_1~(\simeq 62)$, the coefficient of variation is nearly zero; with increasing $D$ in this region, the coefficient of variation increases very slowly. Hence, individual firings of fast spiking interneurons in the case of full synchronization are very regular. However, when passing the 1st threshold $D^*_1$, fast sparse synchronization appears via break-up of full synchronization. Then, the coefficient of variation increases so rapidly, and the irregularity degree of individual firings increases.
In the gray-shaded region, the coefficient of variation continues to increase with relatively slow rates.
Hence, with increasing $D$, spike trains of fast spiking interneurons become more irregular. This increasing tendency for the coefficient of variation continues
in the desynchronized region. For $D=600$ in Fig.~\ref{fig:NSTDP3}(b6), fast spiking interneurons fire more irregularly in comparison with the case of $D=500$, because the value of the coefficient of variation for $D=600$ is increased.

In the case of regular spiking pyramidal cells in the E-population, the coefficients of variation form a well-shaped curve with a minimum at $D \simeq 250$
in Fig.~\ref{fig:NSTDP3}(e). Just after passing $D^*_2$ (e.g., $D=95$), regular spiking pyramidal cells fire very irregularly and sparsely, and hence its coefficient of variation becomes very high. In the states of fast sparse synchronization, the values of coefficient of variation are larger than those in the case of fast spiking interneurons, and hence regular spiking pyramidal cells in the (stimulated) E-population exhibit more irregular spikings than fast spiking interneurons in the (stimulating) I-population. In the case of desynchronization, the increasing tendency in the coefficient of variation continues. However, the increasing rate becomes relatively slow, in comparison with the case of fast spiking interneurons. Thus, for $D=600$, the value of coefficient of variation is less than that
in the case of fast spiking interneurons.

We emphasize that a high coefficient of variation is not necessarily inconsistent with the presence of population synchronous rhythms \cite{W_Review}.
In the gray-shaded region in Figs.~\ref{fig:NSTDP3}(d)-\ref{fig:NSTDP3}(e), fast sparsely synchronized rhythms emerge, together with stochastic and intermittent spike discharges of single cells. Due to the stochastic spike skippings (which results from random phase-lockings to the instantaneous population spike rates), multi-peaked ISI histograms appear. Due to these multi-peaked structure in the histograms, the standard deviation becomes large, which leads to a large coefficient of variation (implying high irregularity). However, in addition to such irregularity, the presence of multi peaks (corresponding to phase lockings) also represents some kind of regularity. In this sense, both irregularity and regularity coexist in spike trains for the case of fast sparse synchronization, in contrast to both cases of full synchronization (complete regularity) and desynchronization (complete irregularity).

We also note that the reciprocal of the coefficient of variation represents regularity degree of individual single-cell spike discharges.
It is expected that high regularity of individual single-cell firings in the $X$-population ($X=$ $I$ or $E$) may result in good population synchronization with high spiking measure $\langle M_s^{(X)} \rangle$ of Eq.~(\ref{eq:SM}). We examine the correlation between the reciprocal of the coefficient of variation and the spiking measure $\langle M_s^{(X)} \rangle$ in both the I- and the E-populations. In the I-population, plots of both the reciprocal of the coefficient of variation and the spiking measure $\langle M_s^{(I)} \rangle$ versus $D$ form downhill-shaped curves, and they are found to have a strong correlation with Pearson's correlation coefficient $r \simeq 0.966$ \cite{Pearson}. On the other hand, in the case of E-population, plots of both the reciprocal of the coefficient of variation and the spiking measure $\langle M_s^{(E)} \rangle$ versus $D$ form bell-shaped curves. They also shows a good correlation, although the Pearson's correlation coefficient is reduced to $r \simeq 0.643$ due to some quantitative discrepancy near $D^*_2$. As a result of such good correlation, the maxima for the reciprocal of the coefficient of variation and the spiking measure $\langle M_s^{(E)} \rangle$ appear at the same value of $D$ ($\simeq 250$).

\subsection{Effect of Interpopulation (both I to E and E to I) STDPs on Population States in The I- and The E-populations}
\label{subsec:ComSTDP}
In this subsection, we consider a combined case including both I to E iSTDP and E to I eSTDP, and study their effects on population states (I, E) by varying the noise intensity $D$ in both the I- and the E-populations. Depending on values of $D$, population-averaged values of saturated interpopulation
synaptic strengths are potentiated (LTP) or depressed (LTD) in comparison to the initial average value, and they make effects on the degree of fast sparse synchronization. Due to the effects of these LTP and LTD, an equalization effect in interpopulation synaptic plasticity is found to emerge in an extended wide range of $D$. In a broad region of intermediate $D$, the degree of good synchronization (with higher synchronization degree) gets decreased due to iLTP (in the case of I to E iSTDP) and eLTD (in the case of E to I eSTDP). On the other hand, in a region of large $D$ the degree of bad synchronization (with lower synchronization degree) becomes increased because of iLTD (in the case of I to E iSTDP) and eLTP (in the case of E to I eSTDP). Particularly, some desynchronized states for $D > D^*_3~(\simeq 537)$ in the absence of STDP becomes transformed into fast sparsely synchronized ones in the presence of interpopulation STDPs, and hence the region of fast sparse synchronization is so much extended. Thus, the degree of fast sparse synchronization becomes nearly the same in such an extended wide region of $D$ (including both the intermediate and the large $D$). We note that this kind of equalization effect is distinctly in contrast to the Matthew (bipolarization) effect in the case of intrapopulation (I to I and E to E) STDPs where good (bad) synchronization becomes better (worse) \cite{SSS,FSS-iSTDP}.

Here, we are concerned about population states (I, E) in the I- and the E-populations for $D > D^*_2~(\simeq 91)$.
In the absence of STDP, (FSS, FSS) appears for $D^*_2 < D < D^*_3~(\simeq 537)$, while for $D>D^*_3$ desynchronization occurs together in both the I- and the E-populations [see Fig.~\ref{fig:NSTDP1}(a)]. The initial synaptic strengths are chosen from the Gaussian distribution with the mean $J_0^{(XY)}$ and the standard deviation $\sigma_0~(=5)$, where $J_0^{(II)}=1300,$ $J_0^{(EE)}=300$, $J_0^{(EI)}=800$, and $J_0^{(IE)}=487.5$ (=$J_0^{(II)} J_0^{(EE)} / J_0^{(EI)}$). (These initial synaptic strengths are the same as those in the absence of STDP.)
We note that this initial case satisfies the E-I ratio balance (i.e., $J_0^{(EE)} / J_0^{(EI)} = J_0^{(IE)} / J_0^{(II)}=0.375$). In the case of combined interpopulation (both I to E and E to I) STDPs, both synaptic strengths $\{ J_{ij}^{(EI)} \}$ and $\{ J_{ij}^{(IE)} \}$ are updated according to the nearest-spike pair-based STDP rule in Eq.~(\ref{eq:MSTDP}), while intrapopulation (I to I and E to E) synaptic strengths are static.
By increasing $D$ from $D^*_2~(\simeq 91)$, we investigate the effects of combined interpopulation STDPs on population states (I, E) in the I- and the E-populations, and make comparison with the case without STDP.

\begin{figure}
\includegraphics[width=0.9\columnwidth]{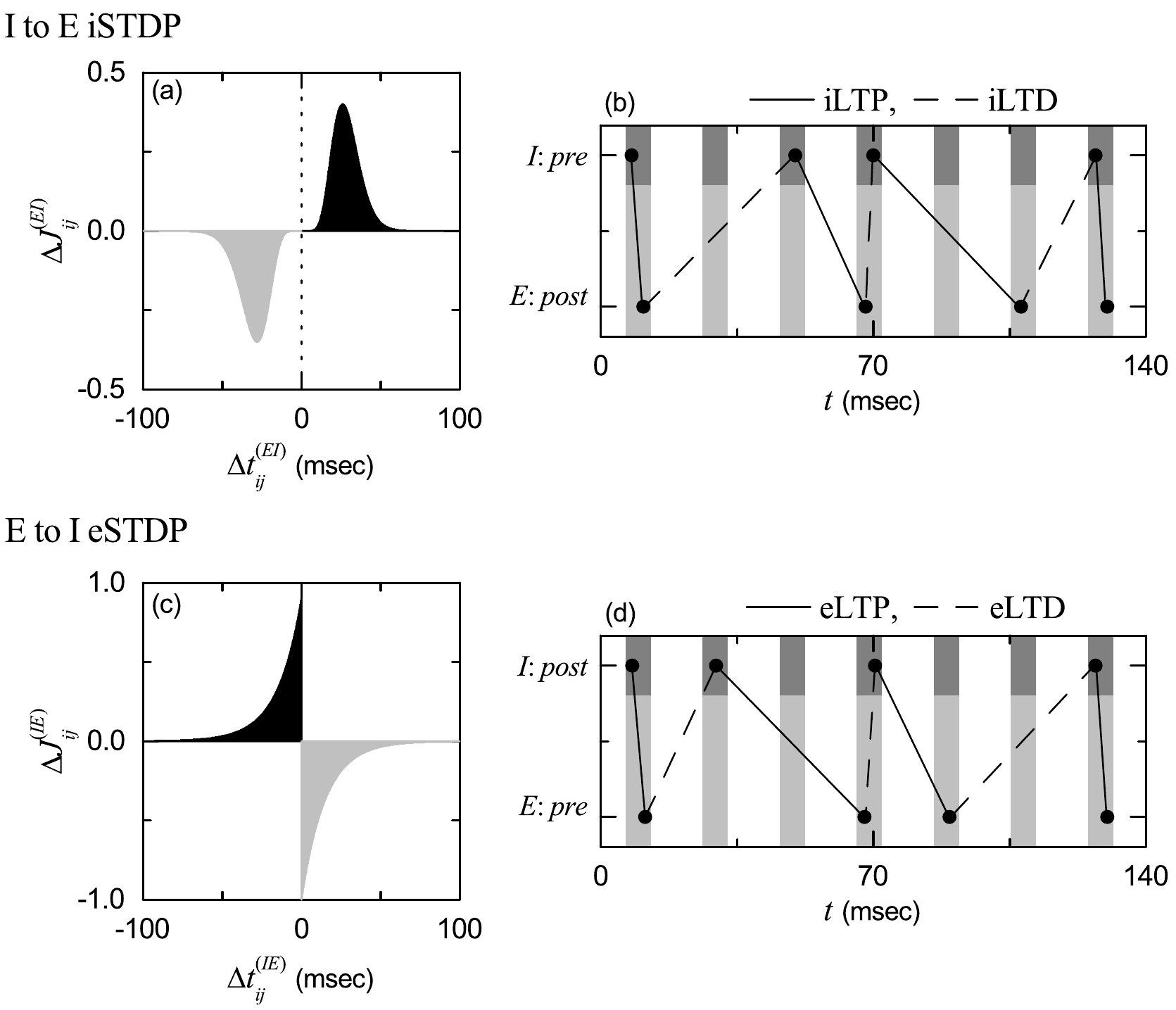}
\caption{Time windows for the interpopulation STDPs and schematic diagrams for the nearest-spike pair-based interpopulation STDP rules.
(a) Time window for the delayed Hebbian I to E iSTDP [see Eq.~(\ref{eq:ItoETW})]. Plot of synaptic modification $\Delta J_{ij}^{(EI)}$  versus
$\Delta t_{ij}^{(EI)}$ $(=t_i^{(post,E)} - t_j^{(pre,I)})$. (b) Schematic diagram for the nearest-spike pair-based I to E iSTDP rule. $I:Pre$ and $E:Post$ correspond to a pre-synaptic fast spiking interneuron and a post-synaptic regular spiking pyramidal cell, respectively.
(c) Time window for the anti-Hebbian E to I eSTDP [see Eq.~(\ref{eq:EtoITW})]. Plot of synaptic modification $\Delta J_{ij}^{(IE)}$ versus $\Delta t_{ij}^{(IE)}$ $(=t_i^{(post,I)} - t_j^{(pre,E)})$. (d) Schematic diagram for the nearest-spike pair-based E to I eSTDP rule. $E:Pre$ and $I:Post$ correspond to a pre-synaptic regular spiking pyramidal cell and a post-synaptic fast spiking interneuron, respectively. 
}
\label{fig:STDP1}
\end{figure}
We first consider the case of I to E iSTDP. Figure \ref{fig:STDP1}(a) shows a time-delayed Hebbian time window for the synaptic modification $\Delta J_{ij}^{(EI)}(\Delta t_{ij}^{(EI)})$ of Eq.~(\ref{eq:ItoETW}) \cite{ItoETW1,ItoETW2,Brazil2}. As in the E to E Hebbian time window \cite{STDP1,STDP2,STDP3,STDP4,STDP5,STDP6,STDP7,STDP8}, LTP occurs in the black region for $\Delta t_{ij}^{(EI)} > 0$, while LTD takes place in the gray region for $\Delta t_{ij}^{(EI)} < 0$. However, unlike the E to E Hebbian time window, $\Delta J_{ij}^{(EI)} \sim 0$ near
$\Delta t_{ij}^{(EI)} \sim 0$, and delayed maximum and minimum for $\Delta J_{ij}^{(EI)}$ appear at
$\Delta t_{ij}^{(EI)} = \beta \tau_+$ and $- \beta \tau_-,$ respectively.

$\Delta J_{ij}^{(EI)}(\Delta t_{ij}^{(EI)})$ varies depending on the relative time difference $\Delta t_{ij}^{(EI)}$ $(=t_i^{(post,E)} - t_j^{(pre,I)})$ between the nearest spike times of the post-synaptic regular spiking pyramidal cell $i$ and the pre-synaptic fast spiking interneuron $j$. When a post-synaptic spike follows a pre-synaptic spike (i.e., $\Delta t_{ij}^{(EI)}$ is positive), inhibitory LTP (iLTP) of I to E synaptic strength appears; otherwise (i.e., $\Delta t_{ij}^{(EI)}$ is negative), inhibitory LTD (iLTD) occurs. A schematic diagram for the nearest-spike pair-based I to E iSTDP rule is given in Fig.~\ref{fig:STDP1}(b), where $I$: $Pre$ and $E$: $Post$ correspond to a pre-synaptic fast spiking interneuron and a post-synaptic regular spiking pyramidal cell, respectively. Here, gray and light gray boxes represent I- and E-stripes in the raster plot of spikes, respectively, and spikes in the stripes are denoted by black solid circles.

When the post-synaptic regular spiking pyramidal cell ($E$: $Post$) fires a spike, iLTP (represented by solid lines) occurs via I to E iSTDP between the post-synaptic spike and the previous nearest pre-synaptic spike of the fast spiking interneuron ($I$: $Pre$). In contrast, when the pre-synaptic
fast spiking interneuron ($I$: $Pre$) fires a spike, iLTD (denoted by dashed lines) occurs through I to E iSTDP between the pre-synaptic spike and the previous nearest post-synaptic spike of the regular spiking pyramidal cell ($E$: $Post$). In the case of fast sparse synchronization, individual neurons make stochastic phase lockings (i.e., they make intermittent spikings phase-locked to the instantaneous population spike rate at random multiples of its global period). As a result of stochastic phase lockings (leading to stochastic spike skippings), nearest-neighboring pre- and post-synaptic spikes may appear in any two separate stripes (e.g., nearest-neighboring, next-nearest-neighboring or farther-separated stripes), as well as in the same stripe, in contrast to the case of full synchronization where they appear in the same or just in the nearest-neighboring stripes [compare Fig.~\ref{fig:STDP1}(b) with Fig. 4(b) (corresponding to the case of full synchronization) in \cite{SSS}]. For simplicity, only the cases, corresponding to the same, the nearest-neighboring, and the next-nearest-neighboring stripes, are shown in Fig.~\ref{fig:STDP1}(b).

Next, we consider the case of E to I eSTDP. Figure \ref{fig:STDP1}(c) shows an anti-Hebbian time window for the synaptic modification
$\Delta J_{ij}^{(IE)}(\Delta t_{ij}^{(IE)})$ of Eq.~(\ref{eq:EtoITW}) \cite{Abbott1,EtoITW1,EtoITW2}.
Unlike the case of the I to E time-delayed Hebbian time window \cite{ItoETW1,ItoETW2,Brazil2}, LTD occurs in the gray region for $\Delta t_{ij}^{(IE)} > 0$, while LTP takes place in the black region for $\Delta t_{ij}^{(IE)} < 0$.
Furthermore, the anti-Hebbian time window for E to I eSTDP is in contrast to the Hebbian time window for the
E to E eSTDP \cite{STDP1,STDP2,STDP3,STDP4,STDP5,STDP6,STDP7,STDP8}, although both cases correspond to the same excitatory synapses (i.e., the type of time window may vary depending on the type of target neurons of excitatory synapses).

The synaptic modification $\Delta J_{ij}^{(IE)}(\Delta t_{ij}^{(IE)})$ changes depending on the relative time difference $\Delta t_{ij}^{(IE)}$ $(=t_i^{(post,I)} - t_j^{(pre,E)})$ between the nearest spike times of the post-synaptic fast spiking interneurons  $i$ and the pre-synaptic regular spiking pyramidal cell $j$. When a post-synaptic spike follows a pre-synaptic spike (i.e., $\Delta t_{ij}^{(IE)}$ is positive), excitatory LTD (eLTD) of E to I synaptic strength occurs; otherwise (i.e., $\Delta t_{ij}^{(IE)}$ is negative), excitatory LTP (eLTP) appears.
A schematic diagram for the nearest-spike pair-based E to I eSTDP rule is given in Fig.~\ref{fig:STDP1}(d), where $E$: $Pre$ and
$I$: $Post$ correspond to a pre-synaptic regular spiking pyramidal cell and a post-synaptic fast spiking interneuron, respectively.
As in the case of I to E iSTDP in Fig.~\ref{fig:STDP1}(b), gray and light gray boxes denote I- and E-stripes in the raster plot, respectively, and spikes in the stripes are represented by black solid circles. When the post-synaptic fast spiking interneuron ($I$: $Post$) fires a spike, eLTD (represented by dashed lines) occurs via E to I eSTDP between the post-synaptic spike and the previous nearest pre-synaptic spike of the regular spiking pyramidal cell ($E$: $Pre$). On the other hand, when the pre-synaptic regular spiking pyramidal cell ($E$: $Pre$) fires a spike, eLTP (denoted by solid lines) occurs through E to I eSTDP between the pre-synaptic spike and the previous nearest post-synaptic spike of the fast spiking interneuron ($I$: $Post$).
In the case of fast sparse synchronization, nearest-neighboring pre- and post-synaptic spikes may appear in any two separate stripes due to stochastic spike skipping. Like the case of I to E iSTDP, only the cases, corresponding to the same, the nearest-neighboring, and the next-nearest-neighboring stripes, are shown in Fig.~\ref{fig:STDP1}(d).

\begin{figure}
\includegraphics[width=0.8\columnwidth]{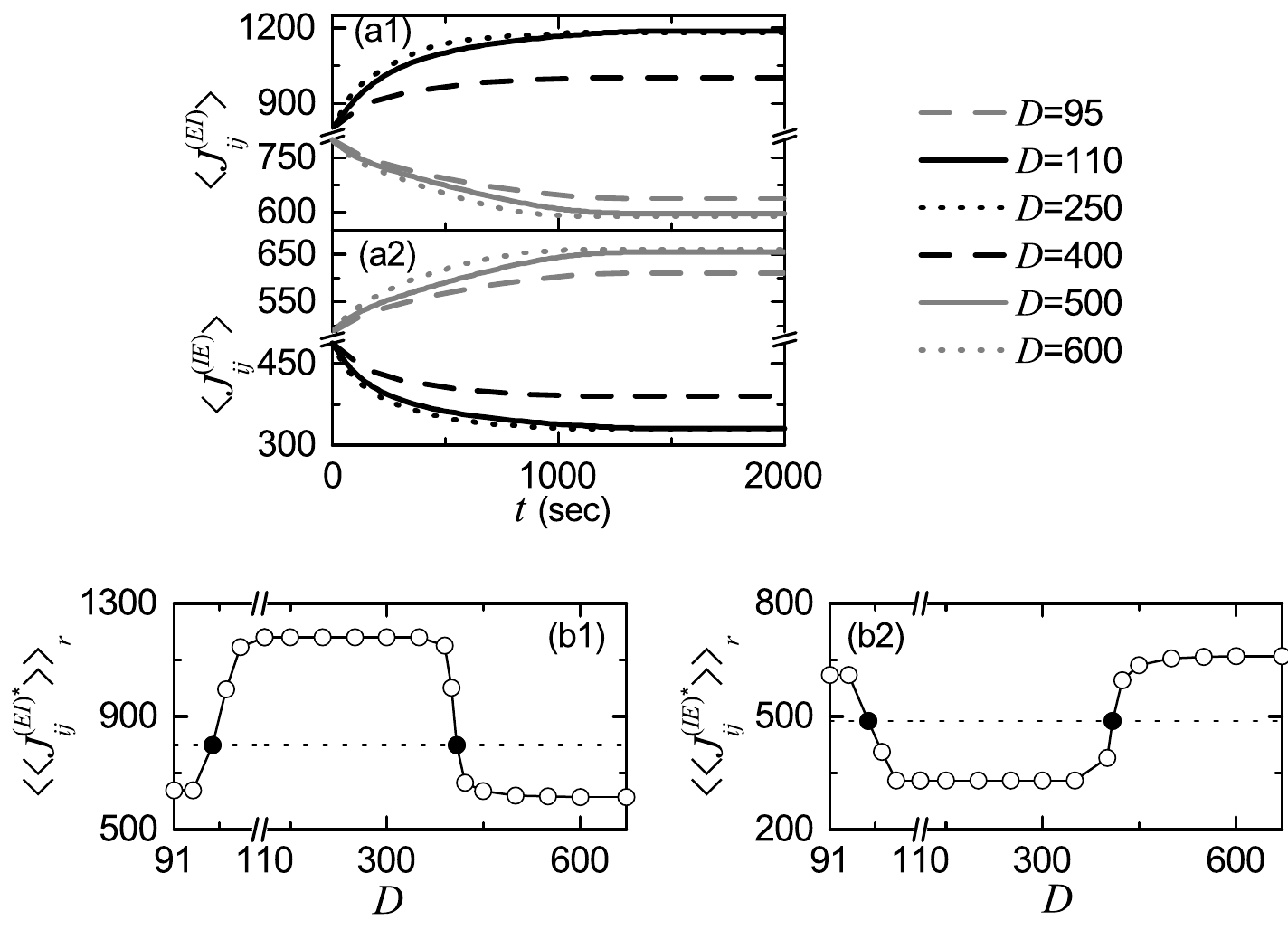}
\caption{Emergence of LTP and LTD in the presence of interpopulation (I to E and E to I) STDPs.
Time-evolutions of population-averaged synaptic strengths (a1) $\langle J_{ij}^{(EI)} \rangle$ and (a2) $\langle J_{ij}^{(IE)} \rangle$ for various values of $D$. Plots of population-averaged saturated limit values of synaptic strengths (b1) $\langle \langle {J_{ij}^{(EI)}}^* \rangle \rangle_r$ and (b2)
$\langle \langle {J_{ij}^{(IE)}}^* \rangle \rangle_r$ versus $D$.
}
\label{fig:STDP2}
\end{figure}

\begin{figure*}
\includegraphics[width=1.7\columnwidth]{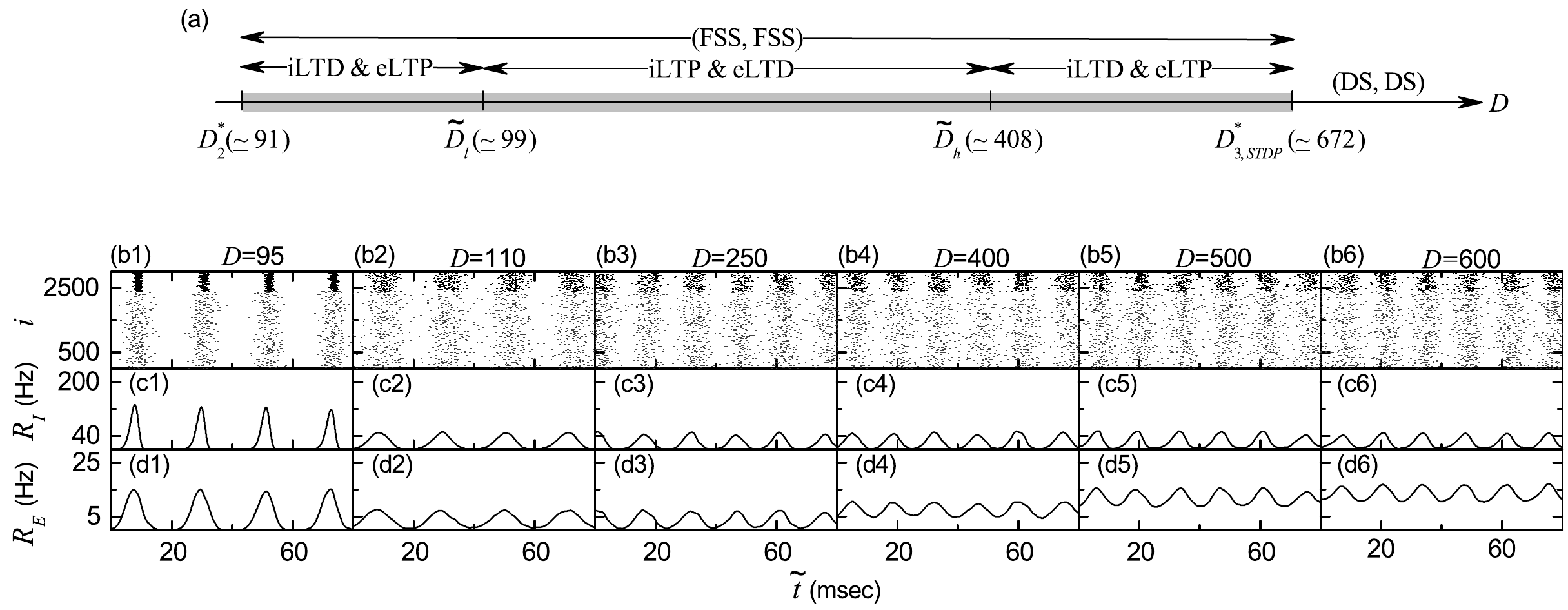}
\caption{Effects of interpopulation (I to E and E to I) STDPs on population states.
(a) Bar diagram for the population states (I, E) in the I- and the E-populations. Raster plots of spikes in (b1)-(b6) and instantaneous population spike rates
$R_I(t)$ in (c1)-(c6) and $R_E(t)$ in (d1)-(d6) for various values of $D$ after the saturation time, where $t=t^*$ (saturation time = 1500 sec) + $\tilde{t}$.
}
\label{fig:STDP3}
\end{figure*}

Figures \ref{fig:STDP2}(a1) and \ref{fig:STDP2}(a2) show time-evolutions of population-averaged I to E synaptic strengths $\langle J_{ij}^{(EI)} \rangle$ and E to I synaptic strengths $\langle J_{ij}^{(IE)} \rangle$ for various values of $D$, respectively.
We first consider the case of $\langle J_{ij}^{(EI)} \rangle$ whose time evolutions are governed by the time-delayed Hebbian time window. In each case of intermediate values of $D=110,$ 250, and 400 (shown in black color), $\langle J_{ij}^{(EI)} \rangle$ increases monotonically above its initial value $J_0^{(EI)}$ (=800), and eventually it approaches a saturated limit value $\langle {J_{ij}^{(EI)}}^* \rangle$ nearly at $t=1500$ sec.
After such a long time of adjustment ($\sim 1500$ sec), the distribution of synaptic strengths becomes nearly stationary (i.e., equilibrated).
Consequently, iLTP occurs for these values of $D$. On the other hand, for small and large values of $D=95,$ 500, and 600 (shown in gray color), $\langle J_{ij}^{(EI)} \rangle$ decreases monotonically below $J_0^{(EI)}$, and approaches a saturated limit value $\langle {J_{ij}^{(EI)}}^* \rangle$. As a result, iLTD occurs in the cases of $D=95,$ 500 and 600.

Next, we consider the case of $\langle J_{ij}^{(IE)} \rangle$. Due to the effect of anti-Hebbian time window, its time evolutions are in contrast to those of $\langle J_{ij}^{(EI)} \rangle$. For intermediate values of $D=110,$ 250, and 400 (shown in black color), $\langle J_{ij}^{(IE)} \rangle$ decreases monotonically below its initial value $J_0^{(IE)}$ (=487.5), and eventually it converges toward a saturated limit value $\langle {J_{ij}^{(IE)}}^* \rangle$ nearly at $t=1500$ sec. As a result, eLTD occurs for these values of $D$. In contrast, for small and large values of $D=95,$ 500, and 600 (shown in gray color), $\langle J_{ij}^{(IE)} \rangle$ increases monotonically above $J_0^{(IE)}$, and converges toward a saturated limit value $\langle {J_{ij}^{(IE)}}^* \rangle$. Consequently, eLTP occurs for $D=95,$ 500 and 600.

Figure \ref{fig:STDP2}(b1) shows a bell-shaped plot of population-averaged saturated limit values $\langle \langle {J_{ij}^{(EI)}}^* \rangle \rangle_r$ (open circles) of I to E synaptic strengths versus $D$ in a range of $D^*_2~(\simeq 91) < D < D^*_{3,STDP}~(\simeq 672)$ where (FSS, FSS) appears in both the I- and the E-populations. Here, the horizontal dotted line represents the initial average value $J_0^{(EI)}$ $(= 800$) of I to E synaptic strengths. In contrast, the plot for population-averaged saturated limit values $\langle \langle {J_{ij}^{(IE)}}^* \rangle \rangle_r$ (open circles) of E to I synaptic strengths  versus $D$ forms a well-shaped graph, as shown in Fig.~\ref{fig:STDP2}(b2), where the horizontal dotted line denotes the initial average value of E to I synaptic strengths $J_0^{(IE)}$ $(= 487.5$). The lower and the higher thresholds, ${\widetilde{D}}_{l}$ ($\simeq 99$) and ${\widetilde{D}}_{h}$ ($\simeq 408$), for LTP/LTD [where $\langle \langle {J_{ij}^{(EI)}}^* \rangle \rangle_r $ and $\langle \langle {J_{ij}^{(IE)}}^* \rangle \rangle_r $ lie on their horizontal lines (i.e., they are the same as
their initial average values, respectively)] are denoted by solid circles. Thus, in the case of a bell-shaped graph for $\langle \langle {J_{ij}^{(EI)}}^* \rangle \rangle_r$, iLTP occurs in a broad region of intermediate $D$ (${\widetilde{D}}_{l} < D < {\widetilde{D}}_{h}$), while iLTD takes place in the other two (separate) regions of small and large $D$ [$D^*_2 < D<{\widetilde{D}}_{l}$ and ${\widetilde{D}}_{h} < D < D^*_{3,STDP}$]. On the other hand, in the case of a well-shaped graph for $\langle \langle {J_{ij}^{(IE)}}^* \rangle \rangle_r$, eLTD takes place in a broad region of intermediate $D$ (${\widetilde{D}}_{l} < D < {\widetilde{D}}_{h}$), while eLTP occurs in the other two (separate) regions of small and large $D$ [$D^*_2 < D<{\widetilde{D}}_{l}$ and ${\widetilde{D}}_{h} < D < D^*_{3,STDP}$].

A bar diagram for the population states (I, E) in the I- and the E-populations is shown in Fig.~\ref{fig:STDP3}(a).
We note that (FSS, FSS) occurs in a broad range of $D$ [$D^*_2~(\simeq 91) < D < D^*_{3,STDP}~(\simeq 672)$],
in comparison with the case without STDP where (FSS, FSS) appears for $D^*_2 < D < D^*_{3}~(\simeq 537)$
[see Fig.~\ref{fig:NSTDP1}(a)]. We note that desynchronized states for $D^*_3 < D < D^*_{3,STDP}$ in the absence of STDP are transformed
into (FSS, FSS) in the presence of combined interpopulation (both I to E and E to I) STDPs, and thus the region of (FSS, FSS) is so much extended.

The effects of LTP and LTD at inhibitory and excitatory synapses on population states after the saturation time ($t^*=1500$ sec) may be well seen in the raster plot of spikes and the corresponding instantaneous population spike rates $R_I(t)$ and $R_E(t)$. Figures \ref{fig:STDP3}(b1)-\ref{fig:STDP3}(b6), Figures \ref{fig:STDP3}(c1)-\ref{fig:STDP3}(c6), and Figures \ref{fig:STDP3}(d1)-\ref{fig:STDP3}(d6) show raster plots of spikes,
the instantaneous population spike rates $R_I(t)$, and the instantaneous population spike rates $R_E(t)$ for various values of $D$, respectively.

In comparison with the case without STDP [see Figs.~\ref{fig:NSTDP1}(b4)-\ref{fig:NSTDP1}(b6), Figs.~\ref{fig:NSTDP1}(c4)-\ref{fig:NSTDP1}(c6), and Figs.~\ref{fig:NSTDP1}(d4)-\ref{fig:NSTDP1}(d6)], the degrees of (FSS, FSS) for intermediate values of $D$ ($D=110,$ 250, and 400) are decreased (i.e., the amplitudes of $R_I(t)$ and $R_E(t)$ are decreased) due to increased I to E synaptic inhibition (iLTP) and decreased E to I synaptic excitation (eLTD).
In contrast, for small and large values of $D$ ($D=95$ and 500), the degrees of (FSS, FSS) are increased (i.e., the amplitudes of $R_I(t)$ and $R_E(t)$ are increased) because of decreased I to E synaptic inhibition (iLTD) and increased E to I synaptic excitation (eLTP) (for comparison, see the corresponding raster plots, $R_I(t)$, and $R_E(t)$ for $D=95$ and 500 in Fig.~\ref{fig:NSTDP1}).
We note that a desynchronized state for $D=600$ in the absence of STDP [see Figs.~\ref{fig:NSTDP1}(b8), \ref{fig:NSTDP1}(c8), and \ref{fig:NSTDP1}(d8)] is transformed into (FSS, FSS) [see Figs.~\ref{fig:STDP3}(b6), \ref{fig:STDP3}(c6), and \ref{fig:STDP3}(d6)] via iLTD and eLTP.
The degree of (FSS, FSS) for $D=600$ is also nearly the same as those for other smaller values of $D$,
because the value of $D=600$ is much far away from its 3rd threshold $D^*_{3,STDP}$ ($\simeq 672$).

\begin{figure}
\includegraphics[width=0.8\columnwidth]{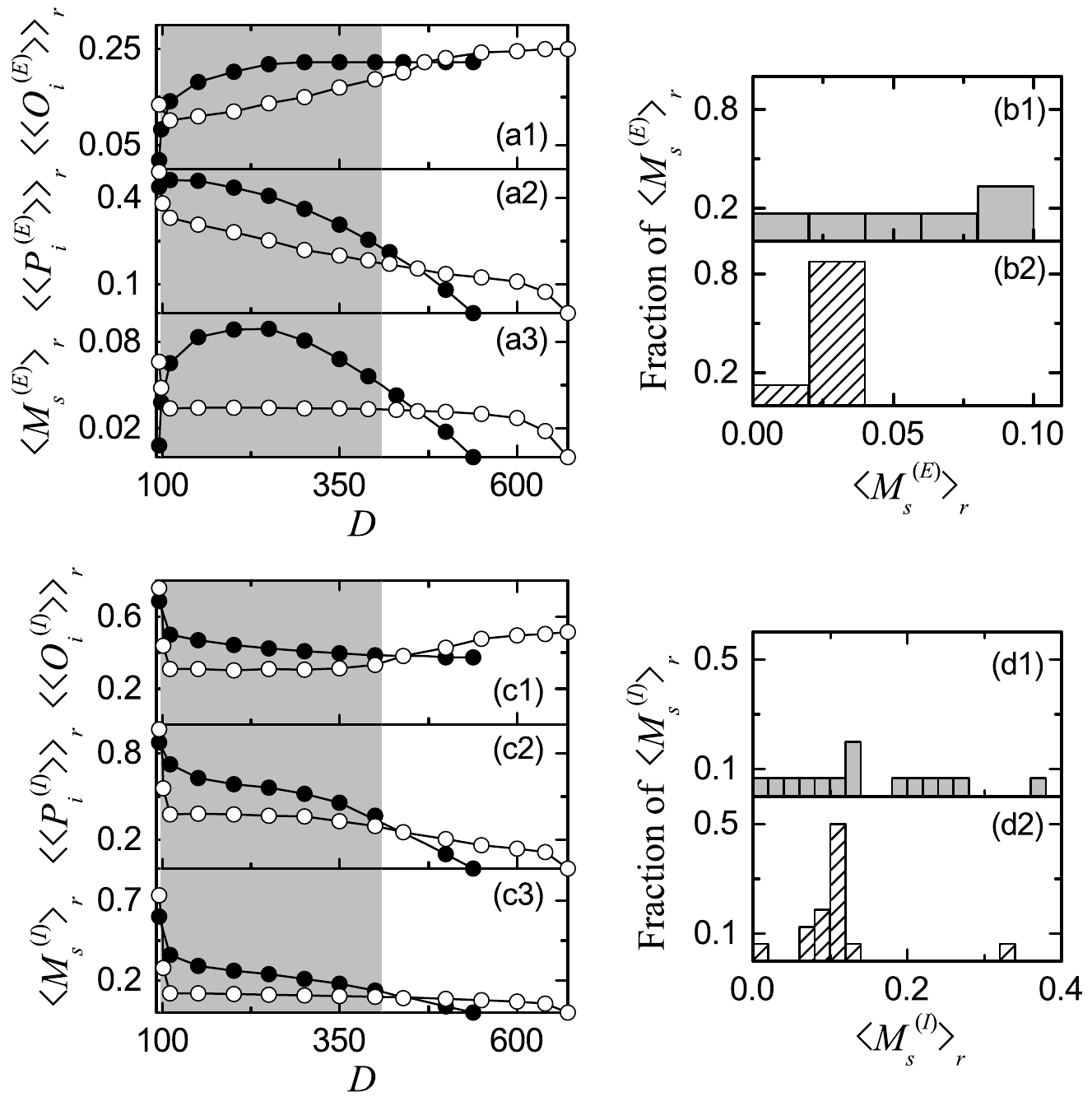}
\caption{Characterization of fast sparse synchronization in the presence of interpopulation (both I to E and E to I) STDPs.
Plots of (a1) the average occupation degree $\langle \langle O_i^{(E)} \rangle \rangle_r$ (open circles), (a2) the average pacing degree $\langle \langle P_i^{(E)} \rangle \rangle_r$  (open circles), and (a3) the statistical-mechanical spiking measure $\langle M_s^{(E)} \rangle_r$  (open circles) versus $D$ in the E-population.
Histograms for distribution of statistical-mechanical spiking measures $\langle M_s^{(E)} \rangle_r$ in the E-population in the (b1) absence and the (b2) presence of
interpopulation (both I to E and E to I) STDPs.
Plots of (c1) the average occupation degree $\langle \langle O_i^{(I)} \rangle \rangle_r$ (open circles), (c2) the average pacing degree $\langle \langle P_i^{(I)} \rangle \rangle_r$ (open circles), and (c3) the statistical-mechanical spiking measure $\langle M_s^{(I)} \rangle_r$  (open circles) versus $D$  in the I-population.
For comparison, $\langle \langle O_i^{(X)} \rangle \rangle_r$, $\langle \langle P_i^{(X)} \rangle \rangle_r$, and $\langle M_s^{(X)}  \rangle_r$  ($X=E$ or $I$) in the absence of STDP are also denoted by solid circles.
Histograms for distribution of statistical-mechanical spiking measures $\langle M_s^{(I)} \rangle_r$ in the I-population in the (d1) absence and the (d2) presence of interpopulation (both I to E and E to I) STDPs.
}
\label{fig:STDP6}
\end{figure}
Here, we also note that the degree of FSS in the I-(E-)population (i.e., the amplitude of $R_I(t)$ [$R_E(t)$]) tends to be nearly the same in an extended wide range of ${\widetilde{D}}_{l} <D < D^*_{3,STDP}$, except for the narrow small-$D$ region ($D^*_2 < D < {\widetilde{D}}_{l}$).
Hence, an equalization effect in the combined interpopulation synaptic plasticity occurs in such an extended wide range of $D$.
Quantitative analysis for the degree of (FSS, FSS) and the equalization effect in the case of combined interpopulation (both I to E and E to I) STDPs will be done intensively in Fig.~\ref{fig:STDP6}.

In an extended wide region of (FSS, FSS) for $D^*_2~(\simeq 91) < D < D^*_{3,STDP}~(\simeq 672)$, we characterize population and individual behaviors for (FSS, FSS) in both the I- and the E-populations, and make comparison with those in the case without STDP.
Population behaviors for fast sparse synchronization in each $X$-population ($X=E$ or $I$) are characterized in terms of the average occupation degree $\langle \langle O_i^{(X)} \rangle \rangle_r$, the average pacing degree $\langle \langle P_i^{(X)} \rangle \rangle_r$, and the statistical-mechanical spiking measure $\langle M_s^{(X)} \rangle_r$. As explained in the subsection \ref{subsec:NSTDP}, $\langle \langle O_i^{(X)} \rangle \rangle_r$ represents average density of spikes in the stripes in the raster plot, $\langle \langle P_i^{(X)} \rangle \rangle_r$ denotes average phase coherence of spikes in the stripes, and $\langle M_s^{(X)} \rangle_r$ (given by a product of occupation and pacing degrees) represents overall degree of fast sparse synchronization.

We first consider the case of E-population which receives I to E synaptic input. Figures \ref{fig:STDP6}(a1)-\ref{fig:STDP6}(a3) show plots of $\langle \langle O_i^{(E)} \rangle \rangle_r,$ $\langle \langle P_i^{(E)} \rangle \rangle_r,$ and $\langle M_s^{(E)} \rangle_r$, respectively.
In the gray region of iLTP [${\widetilde{D}}_{l}~(\simeq 99)  <D < {\widetilde{D}}_{h}~(\simeq 408)]$,
the average occupation degrees $\langle \langle O_i^{(E)} \rangle \rangle_r$ and the average pacing degrees $\langle \langle P_i^{(E)} \rangle \rangle_r$ (open circles) are lower than those (solid circles) in the absence of STDP, due to increased I to E synaptic inhibition. On the other hand, in most cases of iLTD for large $D$ (except for a narrow region near the higher threshold ${\widetilde{D}}_{h}$), $\langle \langle O_i^{(E)} \rangle \rangle_r$ and $\langle \langle P_i^{(E)} \rangle \rangle_r$ (open circles) are higher than those (solid circles) in the absence of STDP, because of decreased I to E synaptic
inhibition.

We are concerned about a broad region of ${\widetilde{D}}_{l}~(\simeq 99)  <D < D^*_{3,STDP}~(\simeq 672)$ (including the regions of both intermediate and large $D$). In this region, $\langle \langle O_i^{(E)} \rangle \rangle_r$ is a relatively fast-increasing function of $D$ (consisting of open circles),
and shows a non-equalization effect, because the standard deviation in the distribution of $\langle \langle O_i^{(E)} \rangle \rangle_r$ is increased in comparison to that in the absence of STDP. In contrast, $\langle \langle P_i^{(E)} \rangle \rangle_r$ is a relatively slowly-decreasing function of $D$
(consisting of open circles) and exhibits a weak equalization effect, because the standard deviation in the distribution of $\langle \langle P_i^{(E)} \rangle \rangle_r$ is decreased in comparison with that in the case without STDP.

The statistical-mechanical spiking measure $\langle M_s^{(E)} \rangle_r$ is given by a product of the occupation and the pacing degrees which exhibit increasing and decreasing behaviors with $D$, respectively.
In the region of intermediate $D$, the degrees of good synchronization (solid circles) in the absence of STDP become decreased to lower ones (open circles) due to
iLTP, while in the region of large $D$ the degrees of bad synchronization (solid circles) in the absence of STDP get increased to higher values (open circles)
because of iLTD. Via the effects of iLTD, even desynchronized states in the absence of STDP are transformed into sparsely synchronized states in the range of $D^*_3~(\simeq 537) < D < D^*_{3,STDP}~(\simeq 672)$, and hence the region of fast sparse synchronization is so much extended in the presence of both I to E and E to I STDPs. In this way, through cooperative interplay between the weak equalization effect in (decreasing) $\langle \langle P_i^{(E)} \rangle \rangle_r$ and the non-equalization effect in (increasing) $\langle \langle O_i^{(E)} \rangle \rangle_r$, strong equalization effect in the spiking measure $\langle M_s^{(E)} \rangle_r$ with much smaller standard deviation is found to occur [i.e., the values of $\langle M_s^{(E)} \rangle_r$ in Fig.~\ref{fig:STDP6}(a3) are nearly the
same]. Thus, a bell-shaped curve (consisting of solid circles) for $\langle M_s^{(E)} \rangle_r$ in the absence of STDP is transformed into a nearly flat curve (composed of open circles) in the presence of combined I to E and E to I STDPs.

This equalization effect may be well seen in the histograms for the distribution of $\langle M_s^{(E)} \rangle_r$.
The gray histogram in the absence of STDP is shown in Fig.~\ref{fig:STDP6}(b1) and the hatched histogram in the presence of
combined I to E and E to I STDPs is given in Fig.~\ref{fig:STDP6}(b2).
The standard deviation ($\simeq 0.007$) in the hatched histogram is much smaller than that ($\simeq 0.028$) in the gray histogram. Hence, strong equalization effect occurs in an extended broad region of ${\widetilde{D}}_{l}~(\simeq 99)  <D < D^*_{3,STDP}~(\simeq 672)$.
We note that this kind of equalization effect is markedly in contrast to the Matthew (bipolarization) effect in the intrapopulation (I to I and E to E) STDPs where good (bad) synchronization gets better (worse) \cite{SSS,FSS-iSTDP}
Furthermore, a dumbing-down effect also occurs because the mean value ($\simeq 0.029$) in the hatched histogram is smaller than that ($\simeq 0.056$) in the gray histogram.

We now consider the case of I-population which receives E to I synaptic input. Figures \ref{fig:STDP6}(c1)-\ref{fig:STDP6}(c3) show plots of $\langle \langle O_i^{(I)} \rangle \rangle_r,$ $\langle \langle P_i^{(I)} \rangle \rangle_r,$ and $\langle M_s^{(I)} \rangle_r$ in the I-population, respectively.
In the gray region of eLTD [${\widetilde{D}}_{l}~(\simeq 99)  <D < {\widetilde{D}}_{h}~(\simeq 408)$],
the average occupation degrees $\langle \langle O_i^{(I)} \rangle \rangle_r$ and the average pacing degrees $\langle \langle P_i^{(I)} \rangle \rangle_r$ (open circles) are lower than those (solid circles) in the absence of STDP, because of decreased E to I synaptic excitation.
In contrast, in most cases of eLTP for large $D$ (except for a narrow region near the higher threshold ${\widetilde{D}}_{h}$), the values of $\langle \langle O_i^{(I)} \rangle \rangle_r$ and $\langle \langle P_i^{(I)} \rangle \rangle_r$ (open circles) are higher than those (solid circles) in the absence of STDP, due to increased E to I synaptic excitation.
In this region of large $D$, $\langle \langle O_i^{(I)} \rangle \rangle_r$ ($\langle \langle P_i^{(I)} \rangle \rangle_r$)
increases (decreases) with $D$ in a relatively fast (slow) way, in contrast to the case without STDP.
Thus, in a wide region of ${\widetilde{D}}_{l}~(\simeq 99)  <D < D^*_{3,STDP}~(\simeq 672)$ (including the regions of both intermediate and large $D$),
the standard deviation in the distribution of $\langle \langle O_i^{(I)} \rangle \rangle_r$ is increased in comparison to that in the absence of STDP, and
thus $\langle \langle O_i^{(I)} \rangle \rangle_r$ exhibits a non-equalization effect. On the other hand, $\langle \langle P_i^{(I)} \rangle \rangle_r$ shows a weak equalization effect, because the standard deviation in the distribution of $\langle \langle P_i^{(I)} \rangle \rangle_r$ is decreased in comparison with that without STDP

The statistical-mechanical spiking measure $\langle M_s^{(I)} \rangle_r$ in the I-population is given by a product of the occupation and the pacing degrees which exhibit increasing and decreasing behaviors with $D$, respectively.
In the region of intermediate $D$, the degrees of good synchronization (solid circles) in the absence of STDP get decreased to lower ones (open circles)
due to eLTD, while in the region of large $D$ the degrees of bad synchronization (solid circles) in the absence of STDP become increased to higher values (open circles) because of eLTP. Through the effects of eLTP, even desynchronized states in the absence of STDP become transformed into sparsely synchronized states in the range of $D^*_3~(\simeq 537) < D < D^*_{3,STDP}~(\simeq 672)$, and hence the region of fast sparse synchronization is so much extended in the presence of combined I to E and E to I STDP. As in the case of $\langle M_s^{(E)} \rangle_r$, via cooperative interplay between the weak equalization effect in (decreasing) $\langle \langle P_i^{(I)} \rangle \rangle_r$ and the non-equalization effect in (increasing) $\langle \langle O_i^{(I)} \rangle \rangle_r$, strong equalization effect in the spiking measure $\langle M_s^{(I)} \rangle_r$ with much smaller standard deviation is found to occur [i.e., the values of $\langle M_s^{(I)} \rangle_r$ in Fig.~\ref{fig:STDP6}(c3) are nearly the same]. This equalization effect in interpopulation synaptic plasticity is distinctly in contrast to the Matthew (bipolarization) effect in the intrapopulation (I to I and E to E) STDPs where good (bad) synchronization gets better (worse) \cite{SSS,FSS-iSTDP}.

This kind of equalization effect may also be well seen in the histograms for the distribution of $\langle M_s^{(I)} \rangle_r$.
The gray histogram in the absence of STDP is shown in Fig.~\ref{fig:STDP6}(d1) and the hatched histogram in the presence of
combined I to E and E to I STDP is given in Fig.~\ref{fig:STDP6}(d2).
The standard deviation ($\simeq 0.056$) in the hatched histogram is much smaller than that ($\simeq 0.112$) in the gray histogram.
Thus, strong equalization effect also occurs in the I-population in an extended broad region of ${\widetilde{D}}_{l}~(\simeq 99)  <D < D^*_{3,STDP}~(\simeq 672)$, as in the case of E-population. Moreover, a dumbing-down effect also occurs because the mean value ($\simeq 0.111$) in the hatched histogram is smaller than that ($\simeq 0.162$) in the gray histogram.

\begin{figure}
\includegraphics[width=\columnwidth]{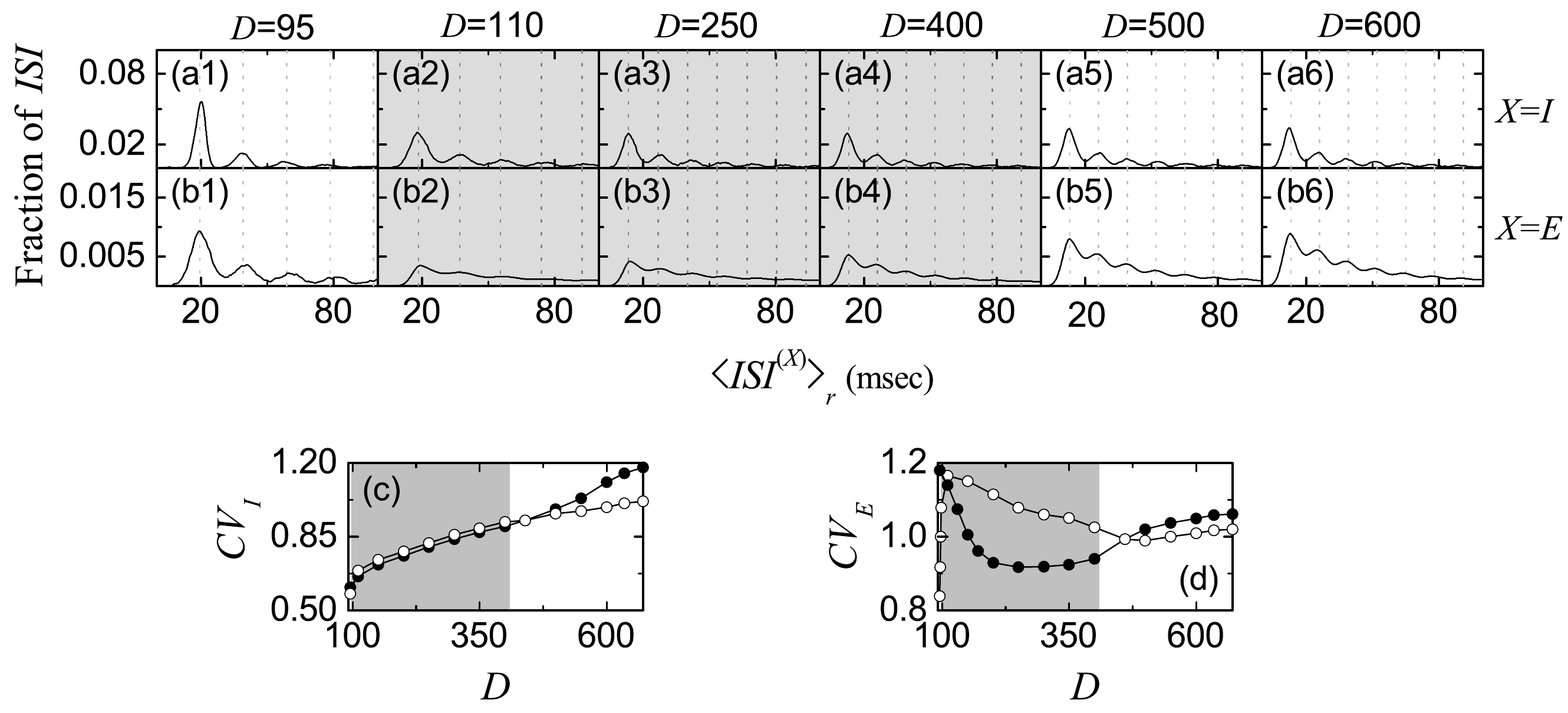}
\caption{
Characterization of individual spiking behaviors in the presence of interpopulation (both I to E and E to I) STDPs. ISI histograms for various values of $D$ in the I-population [(a1)-(a6)] and the E-population [(b1)-(b6)]. Vertical dotted lines in (a1)-(a6) and (b1)-(b6) represent multiples of the global period $T_G^{(X)}$ of the instantaneous population spike rate $R_X(t)$ ($X=I$ or $E$). Plots of the coefficient of variation $CV_X$ (open circles) versus $D$; $X=$ (c) $I$ and (d) $E$.
For comparison, the coefficients of variation in the absence of STDP are represented by solid circles.
}
\label{fig:STDP7}
\end{figure}
From now on, we characterize individual spiking behaviors of fast spiking interneurons and regular spiking pyramidal cells, and compare them with those
in the case without STDP. Figures~\ref{fig:STDP7}(a1)-\ref{fig:STDP7}(a6) [Figures~\ref{fig:STDP7}(b1)-\ref{fig:STDP7}(b6)] show ISI histograms for various values of $D$ in the I-(E-)population. Because of stochastic spike skippings, multiple peaks appear at integer multiples of the global period $T_G^{(X)}$ of $R_X(t)$ ($X=I$ or $E$), as in the case without STDP [see Figs.~\ref{fig:NSTDP3}(b1)-\ref{fig:NSTDP3}(b6) and Figs.~\ref{fig:NSTDP3}(c1)-\ref{fig:NSTDP3}(c6)].
For intermediate values of $D$ (=110, 250, and 400), ISI histograms are shaded in gray color.
In this case of intermediate $D$, iLTP and eLTD occur, and they tend to make single cells fire in a more stochastic and sparse way.
Thus, in these gray-shaded histograms, the 1st-order main peaks become lowered and broadened, higher-order peaks also become wider, and thus mergings between multiple peaks are more developed, when compared with those in the absence of STDP.
Hence, in comparison with those in the case without STDP, the average ISIs $\langle \langle ISI^{(X)} \rangle_r \rangle$
($X=I$ or $E$) become increased, because of the developed tail part. Consequently, population-averaged mean firing rates
$\langle \langle f_i^{(X)} \rangle \rangle_r$ (corresponding to the reciprocals of $\langle \langle ISI^{(X)} \rangle_r \rangle$) are decreased.
These individual spiking behaviors make some effects on population behaviors.
Due to decrease in $\langle \langle f_i^{(X)} \rangle \rangle_r$, spikes become more sparse, and hence the average occupation degree
$\langle \langle O_i^{(X)} \rangle \rangle_r$ in the spiking stripes in the raster plots becomes decreased [see Figs.~\ref{fig:STDP6}(a1) and
\ref{fig:STDP6}(c1)]. Also, because of the enhanced merging between peaks (i.e., due to increase in the irregularity degree
of individual spikings), spiking stripes in the raster plots in Figs.~\ref{fig:STDP3}(b2)-\ref{fig:STDP3}(b4) become more smeared, and hence
the average pacing degrees $\langle \langle P_i^{(X)} \rangle \rangle_r$ of spikes in the stripes get decreased
[see Figs.~\ref{fig:STDP6}(a2) and \ref{fig:STDP6}(c2)].

In contrast, for small and large $D$ (= 95, 500, and 600) iLTD and eLTP occur, and they tend to make individual neurons fire in a less stochastic and sparse way.
Due to the effects of iLTD and eLTP, ISI histograms have much more clear peaks in comparison with
those in the absence of STDP. Particularly, for $D=600$ single-peaked broad ISI histograms in the absence of STDP are transformed into multi-peaked ISI histograms
in the presence of combined I to E and E to I STDPs, because desynchronization in the case without STDP is transformed into fast sparse synchronization in the combined case of both I to E and E to I STDPs.
When compared with those in the absence of STDP, the average ISIs $\langle \langle ISI^{(X)} \rangle_r \rangle$ ($X=I$ or $E$) are decreased due to enhanced lower-order peaks. As a result, population-averaged mean firing rates $\langle \langle f_i^{(X)} \rangle \rangle_r$ are increased.
Because of increase in $\langle \langle f_i^{(X)} \rangle \rangle_r$, the degrees of stochastic spike skipping get decreased, and hence
the average occupation degrees $\langle \langle O_i^{(X)} \rangle \rangle_r$ become increased [see Figs.~\ref{fig:STDP6}(a1) and
\ref{fig:STDP6}(c1)]. Also, due to appearance of clear peaks, spiking stripes in the raster plots in Figs.~\ref{fig:STDP3}(b1), \ref{fig:STDP3}(b5) and \ref{fig:STDP3}(b6) become less smeared, and thus the average pacing degrees $\langle \langle P_i^{(X)} \rangle \rangle_r$ become increased, as shown in
Figs.~\ref{fig:STDP6}(a2) and \ref{fig:STDP6}(c2).

We also study the effects of combined I to E and E to I STDPs on the coefficients of variation (characterizing irregularity degree of individual single-cell firings) in the region of (FSS, FSS) for $D^*_2~(\simeq 91) < D < D^*_{3,STDP}~(\simeq 672)$, and compare them with those in the case without STDP [see Figs.~\ref{fig:NSTDP3}(d) and \ref{fig:NSTDP3}(e)].
Figures \ref{fig:STDP7}(c) and \ref{fig:STDP7}(d) show plots of the coefficients of variations (open circles), $CV_I$ and $CV_E$, for the I- and the E-populations versus $D$, respectively; for comparison, the coefficients of variations (solid circles) in the absence of STDP are also given. Here, the intermediate regions of ${\widetilde{D}}_{l}~ (\simeq 99) < D < {\widetilde{D}}_{h}~(\simeq 408)$ are shaded in gray color.
In the gray-shaded region of intermediate $D$, iLTP and eLTD occur. Then, due to the effects of iLTP and eLTD (tending to increase irregularity degree of single-cell firings), the values of coefficients (open circles) become higher than those (solid circles) in the absence of STDP.
On the other hand, in the other two separate regions of small and large $D$, iLTD and eLTP occur which tend to decrease irregularity degree of single-cell spike discharges. Hence, the values of coefficients (open circles) become lower than those (solid circles) in the absence of STDP, in contrast to the case of intermediate $D$.

As in the case without STDP, we also examine the correlations between the reciprocals of coefficients of variation (denoting the regularity
degree of single-cell spike discharges) and the spiking measures $\langle M_s^{(X)} \rangle_r$ (representing the overall synchronization degree of
fast sparse synchronization) in both the I- and the E-populations ($X=$ $I$ and $E$). Some positive correlations are thus found in the I- and the E-populations
with the Pearson's correlation coefficient $r\simeq 0.858$ and 0.226, respectively.
However, these correlations in the presence of combined STDPs are reduced, in comparison with those in the absence of STDP, mainly due to appearance of strong equalization effects in  $\langle M_s^{(X)} \rangle_r$. In the presence of both I to E and E to I STDPs, standard deviations in the distributions of coefficients of variation are a little decreased, and hence weak equalization effects occur in both the I- and the E-populations, in contrast to strong equalization effects in $\langle M_s^{(X)} \rangle_r$.

\begin{figure}
\includegraphics[width=\columnwidth]{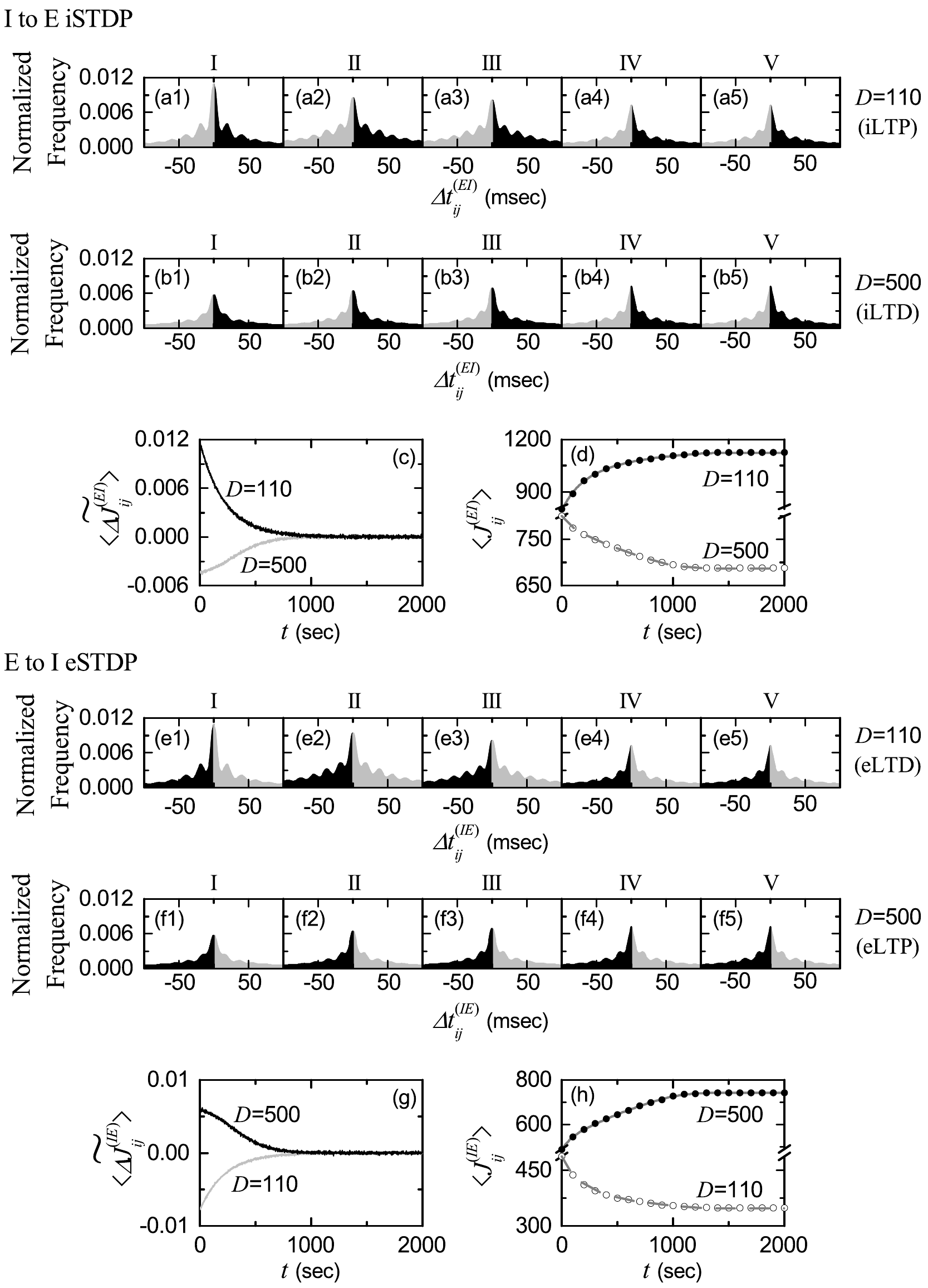}
\caption{Microscopic investigations on emergences of LTP and LTD in the presence of interpopulation (both I to E and E to I) STDPs.
I to E iSTDP: time-evolutions of the normalized histogram $H(\Delta t_{ij}^{(EI)})$  for the distributions of time delays $\{ \Delta t_{ij}^{(EI)} \}$  between the pre- and the post-synaptic spike times for $D=110$ (iLTP) in (a1)-(a5) and for $D=500$ (iLTD) in (b1)-(b5); 5 stages are shown in I (starting from 0 sec), II (starting from 100 sec), III (starting from 400 sec), IV (starting from 800 sec), and V (starting from 1300 sec).
(c) Time-evolutions of multiplicative synaptic modification $\langle {\widetilde{\Delta J_{ij}^{(EI)}}} \rangle$  for $D=110$ (black line) and $D=500$ (gray line). (d) Time-evolutions of population-averaged synaptic strength $\langle J_{ij}^{(EI)} \rangle$ (obtained by an approximate method) for $D=110$ (solid circle) and $D=500$  (open circle).
E to I eSTDP: time-evolutions of the normalized histogram $H(\Delta t_{ij}^{(IE)})$  for the distributions of time delays $\{ \Delta t_{ij}^{(IE)} \}$  between the pre- and the post-synaptic spike times for $D=110$ (eLTD) in (e1)-(e5) and for $D=500$ (eLTP) in (f1)-(f5); 5 stages are shown, as in the above case of I to E iSTDP.
(g) Time-evolutions of multiplicative synaptic modification $\langle {\widetilde{\Delta J_{ij}^{(IE)}}} \rangle$  for $D=110$ (gray line) and $D=500$ (black line). (h) Time-evolutions of population-averaged synaptic strength $\langle J_{ij}^{(IE)} \rangle$ (obtained by an approximate method) for $D=110$ (open circle) and $D=500$ (solid circle).
}
\label{fig:STDP8}
\end{figure}

Finally, we make an intensive investigation on emergences of LTP and LTD (leading to occurrence of equalization effect in interpopulation synaptic plasticity) in the case of combined I to E and E to I STDPs through a microscopic method based on the distributions of time delays $\{ \Delta t_{ij}^{(XY)} \}$ ($=t_i^{(post,X)} - t_j^{(pre,Y)}$)  between the nearest spike times of the post-synaptic neuron $i$ in the $X$-population and the pre-synaptic neuron $j$ in the $Y$-population.
We first consider the case of I to E iSTDP. Figures \ref{fig:STDP8}(a1)-\ref{fig:STDP8}(a5) and Figs.~\ref{fig:STDP8}(b1)-\ref{fig:STDP8}(b5) show time-evolutions of normalized histograms $H(\Delta t_{ij}^{(EI)})$ for the distributions of time delays $\{ \Delta t_{ij}^{(EI)}  \}$ for $D=110$ and 500, respectively; the bin size in each histogram is 0.5 msec. Here, we consider 5 stages, represented by I (starting from 0 sec), II (starting from 100 sec), III (starting from 400 sec), IV (starting from 800 sec), and  V (starting from 1300 sec). At each stage, we obtain the distribution of $\{ \Delta t_{ij}^{(EI)} \}$ for all synaptic pairs during 0.2 sec and get the normalized histogram by dividing the distribution with the total average number of synapses (=96000).

In a case of iLTP for $D=110$, multiple peaks appear in each histogram, which is similar to the case of multi-peaked ISI histogram. As explained in Fig.~\ref{fig:STDP1}(b), due to stochastic spike skippings, nearest-neighboring pre- and post-synaptic spikes appear in any two separate stripes (e.g., nearest-neighboring, next-nearest-neighboring or farther-separated stripes), as well as in the same stripe.
In the stage I, in addition to the sharp main central (1st-order) peak, higher $k$th-order ($k=2,\dots, 5$) left and right minor peaks also are well seen. Here, iLTP and iLTD occur in the black ($\Delta t^{(EI)}>0$) and the gray ($\Delta t^{(EI)} <0$) parts, respectively. As the time $t$ is increased (i.e., with increase in the level of stage), the 1st-order main peak becomes lowered and widened, higher-order peaks also become broadened, and thus mergings between
multiple peaks occur. Thus, at the final stage V, the histogram is composed of lowered and broadened 1st-order peak
and merged higher-order minor peaks. In the stage I, the effect in the right black part (iLTP) is dominant, in comparison with the effect in the left gray part (iLTD), and hence the overall net iLTP begins to emerge. As the level of stage is increased, the effect of iLTP in the black part tends to nearly cancel out the effect of iLTD in the gray part at the stage V.

In a case of iLTD for $D=500$, in the initial stage I, the histogram consists of much lowered and widened 1st-order main peak and higher-order merged peaks, in contrast to the case of $D=110$. For this initial stage, the effect in the left gray part (iLTD) is dominant, in comparison with the effect in the right black part (iLTP), and hence the overall net iLTD begins to occur.
Hence, with increasing the level of stage, the heights of peaks become increased, their widths tend to be narrowed, and thus
peaks (particularly, main peak) become more clear, which is in contrast to the progress in the case of $D=110$.
Moreover, the effect of iLTD in the gray part tends to nearly cancel out the effect of iLTP in the black part at the stage V.
We also note that the two initially-different histograms for both $D=110$ (iLTP) and 500 (iLTD) are evolved into similar ones at the final stage V [see Figs.~\ref{fig:STDP7}(a5) and  \ref{fig:STDP7}(b5)], which shows the equalization effect occurring in the I to E synaptic plasticity.

We consider successive time intervals $I_k \equiv (t_{k},t_{k+1})$, where $t_k=0.2 \cdot (k-1)$ sec ($k=1,2,3,\dots$).
With increasing the time $t$, in each $k$th time interval $I_k$, we get the $k$th normalized histogram
$H_k(\Delta t_{ij}^{(EI)})$ ($k=1,2,3,\dots$) via the distribution of $\{ \Delta t_{ij}^{(EI)} \}$ for all synaptic pairs
during 0.2 sec.
Then, from Eq.~(\ref{eq:MSTDP}), we get the population-averaged synaptic strength $\langle J_{ij}^{(XY)} \rangle_k$ recursively:
\begin{equation}
\langle J_{ij}^{(XY)} \rangle_{k} = \langle J_{ij}^{(XY)} \rangle_{k-1} + \delta \cdot \langle
\widetilde{\Delta J_{ij}^{(XY)}}(\Delta t_{ij}^{(XY)})  \rangle_{k}.
\label{eq:ASS1}
\end{equation}
Here, $X=E$ (post-synaptic population), $Y=I$ (pre-synaptic population), $\langle J_{ij}^{(EI)} \rangle_0=J_0^{(EI)}$ (=800: initial mean value), $\langle \cdots \rangle_k$ in the 2nd term means the average over the distribution of time delays
$\{ \Delta t_{ij}^{(XY)} \}$ for all synaptic pairs in the $k$th time interval, and the multiplicative synaptic modification $\widetilde{\Delta J_{ij}^{(XY)}}(\Delta t_{ij}^{(XY)})$ is given
by the product of the multiplicative factor ($J^*-J_{ij}^{(XY)}$) [$J_{ij}^{(XY)}:$ synaptic coupling strength at the $(k-1)$th stage] and the absolute value of synaptic modification $| \Delta J_{ij}^{(XY)}(\Delta t_{ij}^{(XY)}) |$:
\begin{equation}
 \widetilde{\Delta J_{ij}^{(XY)}}(\Delta t_{ij}^{(XY)})  =  (J^* - J_{ij}^{(XY)})~ |\Delta J_{ij}^{(XY)}(\Delta t_{ij}^{(XY)})|.
\label{eq:ASS2}
\end{equation}
Here, we obtain the population-averaged multiplicative synaptic modification
$\langle \widetilde{\Delta J_{ij}^{(XY)}}(\Delta t_{ij}^{(XY)}) \rangle_{k}$ for the $k$th stage
through a population-average approximation where $J_{ij}^{(XY)}$ is replaced by its population average
$\langle J_{ij}^{(XY)} \rangle_{k-1}$ at the $(k-1)$th stage:
\begin{widetext}
\begin{equation}
 \langle \widetilde{\Delta J_{ij}^{(XY)}}(\Delta t_{ij}^{(XY)}) \rangle_k  \simeq (J^*- \langle J_{ij}^{(XY)} \rangle_{k-1})~ \langle |\Delta J_{ij}^{(XY)}(\Delta t_{ij}^{(XY)})| \rangle_k.
\label{eq:ASS3}
\end{equation}
Here, $\langle |\Delta J_{ij}^{(XY)}(\Delta t_{ij}^{(XY)})| \rangle_k$ may be easily got from the $k$th normalized histogram $H_k(\Delta t_{ij}^{(XY)})$:
\begin{equation}
  \langle |\Delta J_{ij}^{(XY)}(\Delta t_{ij}^{(XY)})| \rangle_{k}  \simeq  \sum_{\rm bins} H_{k} (\Delta t_{ij}^{(XY)}) \cdot | \Delta J_{ij}^{(XY)} (\Delta t_{ij}^{(XY)}) |.
\label{eq:ASS4}
\end{equation}
\end{widetext}
Using Eqs.~(\ref{eq:ASS1}), (\ref{eq:ASS3}), and (\ref{eq:ASS4}), we get approximate values of $\langle \widetilde{\Delta J_{ij}^{(XY)}} \rangle_k$ and $\langle J_{ij}^{(XY)} \rangle_{k}$ in a recursive way.

Figure \ref{fig:STDP8}(c) shows time-evolutions of $\langle \widetilde{\Delta J_{ij}^{(EI)}} \rangle$ for $D=110$ (black curve) and $D=500$ (gray curve).
$\langle \widetilde{\Delta J_{ij}^{(EI)}} \rangle$ for $D=110$ is positive, while $\langle \widetilde{\Delta J_{ij}^{(EI)}} \rangle$ for $D=500$ is negative. For both cases they converge toward nearly zero at the stage V (starting from 1300 sec) because the effects of iLTP and iLTD in the normalized histograms are nearly cancelled out. The time-evolutions of $\langle J_{ij}^{(EI)} \rangle$ for $D=110$ (solid circles) and $D=500$ (open circles) are also shown in Fig.~\ref{fig:STDP8}(d). We note that the approximately-obtained values for $\langle J_{ij}^{(EI)} \rangle$ agree well with directly-obtained ones [denoted by the gray solid (dashed) line for $D=110$ (500)] in Fig.~\ref{fig:STDP2}(a1).
As a result, iLTP (iLTD) emerges for $D=110$ (500).

As in the case of I to E iSTDP, we now study emergences of eLTD and eLTP in E to I eSTDP
via a microscopic method based on the distributions of time delays $\{ \Delta t_{ij}^{(IE)} \}$
($=t_i^{(post,I)} - t_j^{(pre,E)}$) between the nearest spike times of the post-synaptic fast spiking interneuron $i$ and
the pre-synaptic regular spiking pyramidal cell $j$. Figures \ref{fig:STDP8}(e1)-\ref{fig:STDP8}(e5) and Figs.~\ref{fig:STDP8}(f1)-\ref{fig:STDP8}(f5) show
time-evolutions of normalized histograms $H(\Delta t_{ij}^{(IE)})$ for the distributions of time delays
$\{ \Delta t_{ij}^{(IE)} \}$ for $D=110$ and 500, respectively; the bin size in each histogram is 0.5 msec.
Here, we also consider 5 stages, as in the case of the above I to E iSTDP.
At each stage, we get the distribution of $\{ \Delta t_{ij}^{(IE)} \}$ for all synaptic pairs during 0.2 sec and obtain the normalized histogram by dividing the distribution with the total average number of synapses (=96000).

As an example of eLTD in the region of intermediate $D$, we consider the case of $D=110$.
Due to stochastic spike skippings for fast spike skippings, multiple peaks appear in each histogram, as in the multi-peaked ISI histograms.
In the stage I, along with the sharp main central (1st-order) peak, higher $k$th-order ($k=2,\dots, 5$) left and right minor peaks also are well seen. Because of the anti-Hebbian time window for the E to I eSTDP, eLTD and eLTP occur in the gray ($\Delta t^{(IE)}>0$) and the black ($\Delta t^{(IE)} <0$) parts, respectively, which is in contrast to the case of I to E iSTDP with a time-delayed Hebbian time window where iLTP and iLTD occur in the black ($\Delta t^{(IE)}>0$) and the gray ($\Delta t^{(IE)} <0$) parts, respectively [see Figs.~\ref{fig:STDP1}(a) and \ref{fig:STDP1}(c)].
With increasing the level of stage, the 1st-order main peak becomes lowered and broadened, higher-order peaks also become widened, and thus mergings between multiple peaks occur. Thus, at the final stage V, the histogram consists of lowered and widened 1st-order peak and merged higher-order minor peaks.
In the stage I, the effect in the right gray part (eLTD) is dominant, in comparison to the effect in the left black part (eLTP), and hence the overall net eLTD begins to appear. As the level of stage is increased, the effect of eLTD in the gray part tends to nearly cancel out the effect of eLTP in the black part at the stage V.

We consider another case of $D=500$ where eLTP occurs.
In the initial stage I, the histogram is composed of much lowered and broadened 1st-order main peak and higher-order merged peaks, in contrast to the case of $D=110$. For this initial stage, the effect in the left black part (eLTP) is dominant, when compared with the effect in the right gray part (eLTD), and hence the overall net eLTP begins to occur.
Hence, as the level of stage is increased, the heights of peaks become increased, their widths tend to be narrowed, and thus
peaks become more clear, in contrast to the progress in the case of $D=110$.
Furthermore, the effect of eLTP in the black part tends to nearly cancel out the effect of eLTD in the gray part at the stage V.
We also note that the two initially-different histograms in the cases of $D=110$ and 500 are developed into similar ones at the final stage V [see Figs.~\ref{fig:STDP8}(e5) and  \ref{fig:STDP8}(f5)], which shows the equalization effect occurring in the case of E to I eSTDP.

As in the case of I to E iSTDP, we consider successive time intervals $I_k \equiv (t_{k},t_{k+1})$, where $t_k=0.2 \cdot (k-1)$ sec ($k=1,2,3,\dots$). As the time $t$ is increased, in each $k$th time interval $I_k$, we obtain the $k$th normalized histogram
$H_k(\Delta t_{ij}^{(IE)})$ ($k=1,2,3,\dots$) through the distribution of $\{ \Delta t_{ij}^{(IE)} \}$ for all synaptic pairs
during 0.2 sec. Then, using Eqs.~(\ref{eq:ASS1}), (\ref{eq:ASS3}), and (\ref{eq:ASS4}), we obtain approximate values of
multiplicative synaptic modification $\langle \widetilde{\Delta J_{ij}^{(IE)}} \rangle_k$ and population-averaged synaptic strength $\langle J_{ij}^{(IE)} \rangle_{k}$ in a recursive way.
Figure \ref{fig:STDP8}(g) shows time-evolutions of $\langle \widetilde{\Delta J_{ij}^{(IE)}} \rangle$ for $D=110$ (gray curve) and $D=500$ (black curve). $\langle \widetilde{\Delta J_{ij}^{(IE)}} \rangle$ for $D=110$ is negative, while $\langle \widetilde{\Delta J_{ij}^{(IE)}} \rangle$ for $D=500$ is positive. For both cases they converge toward nearly zero at the stage V (starting from 1300 sec) because the effects of eLTP and eLTD in the normalized histograms are nearly cancelled out.
The time-evolutions of $\langle J_{ij}^{(IE)} \rangle$ for $D=110$ (open circles) and $D=500$ (solid circles) are also shown in Fig.~\ref{fig:STDP8}(h). We note that the approximately-obtained values for $\langle J_{ij}^{(IE)} \rangle$ agree well with directly-obtained ones [represented by the gray dashed (solid) line for $D=110$ (500)] in Fig.~\ref{fig:STDP2}(a2).
Consequently, eLTD (eLTP) emerges for $D=110$ (500), in contrast to the case of I to E iSTDP where
iLTP (iLTD) occurs for $D=110$ (500).

\section{Summary and Discussion}
\label{sec:SUM}
We are interested in fast sparsely synchronized brain rhythms, related to diverse cognitive functions such as feature integration, selective attention, and memory formation \cite{W_Review}. In most cases of previous works, emergence of fast sparsely synchronized rhythms and their properties have been studied for static synaptic strengths (i.e., without considering synaptic plasticity) in single-population networks of purely inhibitory interneurons and in two-population networks composed of inhibitory interneurons and excitatory pyramidal cells \cite{Sparse1,Sparse2,Sparse3,Sparse4,Sparse5,Sparse6}. Only in one case \cite{FSS-iSTDP}, intrapopulation I to I iSTDP was considered in an inhibitory small-world network of fast spiking interneurons. In contrast to these previous works, in the present work, we took into consideration adaptive dynamics of interpopulation (both I to E and E to I) synaptic strengths, governed by the I to E iSTDP and the E to I eSTDP, respectively. We also note that fast sparsely synchronized rhythms appear, independently of network structure. Here, we considered clustered small-world networks with both I- and E-populations. The inhibitory small-world network is composed of fast spiking interneurons, and the excitatory small-world network consists of regular spiking pyramidal cells. A time-delayed Hebbian time window has been used for the I to E iSTDP update rule, while an anti-Hebbian time window has been employed for the E to I eSTDP update rule.

By varying the noise intensity $D,$ we have investigated the effects of interpopulation STDPs on diverse population and individual properties of fast sparsely synchronized rhythms that emerge in both the I- and the E-populations in the combined case of both I to E and E to I STDPs. In the presence of interpopulation STDPs, the distribution of interpopulation synaptic strengths $\{ J_{ij}^{(XY)} \}$ is evolved into a saturated one after a sufficiently long time of adjustment. Depending on $D$, the mean $\langle \langle {J_{ij}^{(XY)}}^* \rangle \rangle_{r}$ for saturated limit interpopulation synaptic strengths has been found to increase or decrease [i.e., emergence of LTP or LTD]. These LTP and LTD make effects on the degree of fast sparse synchronization.

In the case of I to E iSTDP, increase (decrease) in the mean $\langle \langle {J_{ij}^{(EI)}}^* \rangle \rangle_r$ of the I to E synaptic inhibition
has been found to disfavor (favor) fast sparse synchronization [i.e. iLTP (iLTD) tends to decrease (increase) the degree of fast sparse synchronization].
In contrast, the roles of LTP and LTD are reversed in the case of E to I eSTDP. In this case, eLTP (eLTD) in the E to I synaptic excitation has been found to
favor (disfavor) fast sparse synchronization [i.e.,  increase (decrease) in the mean $\langle \langle {J_{ij}^{(IE)}}^* \rangle \rangle_r$ tends to increase (decrease) the degree of fast sparse synchronization].
Particularly, desynchronized states for $D^*_3~(\simeq 537)~<D<~ D^*_{3,STDP}~(\simeq 672)$ in the absence of STDP become transformed into fast sparsely
synchronized ones via iLTD and eLTP in the presence of interpopulation STDPs, and hence the region of fast sparse synchronization is so much extended.

An equalization effect in interpopulation (both I to E and E to I) synaptic plasticity has been found to occur in
such an extended wide range of $D$. In a broad region of intermediate $D$, the degree of good synchronization (with higher synchronization degree) gets decreased due to iLTP (in the case of I to E iSTDP) and eLTD (in the case of E to I eSTDP). On the other hand, in a region of large $D$, the degree of bad synchronization (with lower synchronization degree) becomes increased because of iLTD (in the case of I to E iSTDP) and eLTP (in the case of E to I eSTDP). As a result, the degree of fast sparse synchronization in each E- or I-population becomes nearly the same in a wide range of $D$.
Due to the equalization effect in interpopulation STDPs, fast sparsely synchronized rhythms seem to be evolved into stable and robust ones (i.e., less sensitive ones) against external noise in an extended wide region (including both the intermediate and the large $D$).

This kind of equalization effect in interpopulation synaptic plasticity is markedly in contrast to the Matthew (bipolarization) effect in intrapopulation (I to I and E to E) synaptic plasticity where good (bad) synchronization becomes better (worse) \cite{SSS,FSS-iSTDP}.
In this case of Matthew effect, in a broad region of intermediate $D$, fast sparsely synchronized rhythms are so much enhanced due to a constructive effect of intrapopulation STDPs to increase their synchronization degrees, while in a region of large $D$ they are so much depressed or they are transformed into desynchronized states due to a destructive effect of intrapopulation STDPs to decrease their synchronization degrees.
Thus, in the presence of intrapopulation STDPs, the better good synchronization (with higher synchronization degree) becomes, the worse bad synchronization (with lower synchronization degree) gets (e.g., see Fig.~6 in \cite{SSS} and Fig.~7 in \cite{FSS-iSTDP}). This type of Matthew (bipolarization) effect may be observed in diverse aspects in many fields \cite{ME1,ME2}, and it is sometimes summarized by an adage, ``the rich get richer and the poor get poorer'' (representing bipolarization).

We note that the spiking measure $\langle M_s^{(X)} \rangle_r$ (denoting the synchronization degree) is given by the product of the occupation (representing density of spiking stripes) and the pacing (denoting phase coherence between spikes) degrees of spikes in the raster plot. Due to interpopulation STDPs, the average pacing degree $\langle \langle P_i^{(X)} \rangle \rangle_r$ has been found to exhibit a kind of weak equalization effect (i.e., $\langle \langle P_i^{(X)} \rangle \rangle_r$ is a relatively slowly-decreasing function of $D$ with a smaller standard deviation, in comparison with $\langle \langle P_i^{(X)} \rangle \rangle_r$  in the absence of STDP). On the other hand, the average occupation degree $\langle \langle O_i^{(X)} \rangle \rangle_r$ has been found to show a type of non-equalization effect (i.e., $\langle \langle O_i^{(X)} \rangle \rangle_r$ is an increasing function of $D$ with a larger standard deviation, when compared with $\langle \langle O_i^{(X)} \rangle \rangle_r$ in the absence of STDP). Through cooperative interplay between the weak equalization effect in (decreasing) $\langle \langle P_i^{(X)} \rangle \rangle_r$ and the non-equalization effect in (increasing) $\langle \langle O_i^{(X)} \rangle \rangle_r$, strong equalization effect in $\langle M_s^{(X)} \rangle_r$ with much smaller standard deviation has been found to emerge (i.e., the curve for $\langle M_s^{(X)} \rangle_r$ becomes nearly flat in a wide range of $D$).

This kind of equalization effect can be well visualized in the histograms for the spiking measures $\langle M_s^{(X)} \rangle_r$ in the presence and in the absence of interpopulation STDPs. In each E- or I-population, the standard deviation from the mean in the histogram in the case of interpopulation STDPs has been found to be much smaller than that in the case without STDP, which clearly shows emergence of the equalization effect in both the I- and the E-populations. Moreover, a dumbing-down effect in interpopulation synaptic plasticity has also been found to occur in both the E- and the I-populations, because the mean in the histogram in the case of interpopulation STDPs is smaller than that in the absence of STDP. Thus, in each E- or I-population, equalization effect occurs together with dumbing-down effect.

In addition to the above population behaviors, effects of combined I to E and E to I STDPs on individual spiking behaviors have also been investigated.
In the case of I to E iSTDP, iLTP (iLTD) has been found to increase (decrease) irregularity degree of individual single-cell spike discharges
[i.e., due to iLTP (iLTD), single cells tend to fire more (less) irregularly and sparsely].
On the other hand, in the case of E to I eSTDP, the roles of LTP and LTD are reversed. Hence, irregularity degree of individual single-cell firings
has been found to decrease (increase) due to eLTP (eLTD) [i.e., because of eLTP (eLTD), single cells tend to make firings less (more) irregularly and sparsely.

LTP and LTD in both I to E and E to I STDPs make effects on distributions of ISIs.
In the case of fast sparse synchronization, multi-peaked ISI histograms appear due to stochastic spike skippings, in contrast to the case of full synchronization with a single-peaked ISI histogram. In the region of intermediate $D$, due to the effects of iLTP and eLTD, the 1st-order main peaks become lowered and widened, higher-order peaks also become broader, and thus mergings between multiple peaks are more developed, in comparison with those in the absence of STDP.
Thus, the average ISIs $\langle \langle ISI^{(X)} \rangle_r \rangle$ ($X=I$ or $E$) become increased, because of the developed tail part.
As a result, population-averaged mean firing rates $\langle \langle f_i^{(X)} \rangle \rangle_r$ (corresponding to the reciprocals of $\langle \langle ISI^{(X)} \rangle_r \rangle$) get decreased. On the other hand, in the region of small and large $D$, because of the effects of iLTD and eLTP, ISI histograms have much more clear peaks when compared with those in the absence of STDP. As a result, the average ISIs $\langle \langle ISI^{(X)} \rangle_r \rangle$ ($X=I$ or $E$) become decreased due to enhanced lower-order peaks. Consequently, population-averaged mean firing rates $\langle \langle f_i^{(X)} \rangle \rangle_r$ get increased.

Furthermore, effects of both I to E and E to I STDPs on the coefficients of variation (characterizing the irregularity degree of individual single-cell spike discharge) have also been studied in both the I- and the E-populations. In the intermediate region where iLTP and eLTD occurs, irregularity degrees of individual single-cell firings increase, and hence the coefficients of variation become increased. In contrast, in the other two separate regions of small and large $D$ where iLTD and eLTP occur, the degrees of irregularity of individual spikings decrease, and hence the coefficients of variation get decreased.
Reciprocals of coefficients of variation (representing the regularity degree of individual single-cell firings) have also been found to have positive correlations with the spiking measures (denoting the overall synchronization degree of fast sparse synchronization).

Particularly, emergences of LTP and LTD of interpopulation synaptic strengths (leading to occurrence of equalization effect in interpopulation synaptic plasticity) were investigated via a microscopic method, based on the distributions of time delays $\{ \Delta t_{ij}^{(XY)} \}$ between the nearest spiking times of the post-synaptic neuron $i$ in the (target) $X$-population and the pre-synaptic neuron $j$ in the (source) $Y$-population.
Time evolutions of normalized histograms $H(\Delta t_{ij}^{(XY)})$ were followed in both cases of LTP and LTD.
We note that, due to the equalization effects, the normalized histograms (in both cases of LTP and LTD) at the final
(evolution) stage are nearly the same, which is in contrast to the cases of intrapopulation (I to I and E to E) STDPs where the two normalized histograms at the final stage are distinctly different because of the Matthew (bipolarization) effect (e.g., see Fig.~8 in \cite{SSS} and Fig.~8 in \cite{FSS-iSTDP}). Employing a recurrence relation, we recursively obtained population-averaged interpopulation synaptic strength $\langle J_{ij}^{(XY)} \rangle$ at successive stages via an approximate calculation of population-averaged multiplicative synaptic modification $\langle \widetilde{\Delta J_{ij}^{(XY)}} \rangle$ of Eq.~(\ref{eq:ASS3}), based on the normalized histogram at each stage. These approximate values of $\langle J_{ij}^{(XY)} \rangle$ have been found to agree well with directly-calculated ones. Consequently, one can understand clearly how microscopic distributions of $\{ \Delta t_{ij}^{(XY)} \}$ contribute to $\langle J_{ij}^{(XY)} \rangle$.

Finally, we discuss limitations of our present work and future works.
In the present work, we have restricted out attention just to interpopulation (I to E and E to I) STDPs and found emergence
of equalization effects. In previous works, intrapopulation (I to I and E to E) STDPs were studied and the Matthew (bipolarization) effects were found to appear \cite{SSS,FSS-iSTDP}. Hence, in future work, it would be interesting to study competitive interplay between the equalization effect in interpopulation synaptic plasticity and the Matthew (bipolarization) effect in intrapopulation synaptic plasticity in networks consisting of both E- and I-populations with both intrapopulation and interpopulation STDPs. In addition to fast sparsely synchronized rhythms (main concern in the present study), asynchronous irregular states (which show stationary global activity and stochastic sparse spike discharges of single cells) also appear in the hippocampal and the neocortical networks.
Therefore, as another future work, it would be interesting to study mechanisms (provided by the STDPs in networks with both E- and I-populations) for emergence of asynchronous irregular states which are known to play an important role of information processing \cite{STDP1,iSTDP6}. Fast sparsely synchronized rhythms appear, independently of network structure. Here, we considered clustered small-world networks. It is expected that the effects of interpopulation STDPs would also be
independent of network architecture, which will be examined in a future work.

\section*{Acknowledgments}
This research was supported by the Basic Science Research Program through the National Research Foundation of Korea (NRF) funded by the Ministry of Education (Grant No. 20162007688).

\end{document}